\documentclass{article}

\usepackage[english]{babel}

\usepackage[letterpaper,top=2cm,bottom=2cm,left=3cm,right=3cm,marginparwidth=1.75cm]{geometry}

\usepackage{amsmath}
\usepackage{graphicx}
\usepackage[colorlinks=true, allcolors=blue]{hyperref}
\usepackage[utf8]{inputenc} 
\usepackage[T1]{fontenc}    
\usepackage{natbib}
\usepackage{hyperref}       
\usepackage{url}            
\usepackage{booktabs}       
\usepackage{amsfonts}       
\usepackage{nicefrac}       
\usepackage{microtype}      
\usepackage{lipsum}
\usepackage{fancyhdr}       
\usepackage{graphicx}       
\graphicspath{{media/}}     
  
\usepackage{authblk}

\title{Nested MCMC for Uncertainty Propagation in Spatial Modeling of Temporal Long-Memory in Resting-State fMRI}
\date{}
\author[1]{Yasaman Shahhosseini}
\author[2]{C\'edric Beaulac}
\author[1]{Farouk S. Nathoo}
\author[1]{Michelle F. Miranda}

\affil[1]{Department of Mathematics and Statistics, University of Victoria, Victoria, BC, Canada}
\affil[2]{D\'epartement de math\'ematiques, Universit\'e du Qu\'ebec \`a Montr\'eal, Montr\'eal, QC, Canada}

\begin{document}
\maketitle

\begin{abstract}
Resting-state functional magnetic resonance imaging (rs-fMRI) signals exhibit long-range temporal dependence, characterized by a slow power-law decay of autocorrelation. This scale-free temporal structure, quantified by the long-memory (LM) parameter, has been associated with brain function and age-related changes in neural dynamics. Investigating spatial variation in cortical LM and its association with demographic and clinical covariates requires methods that account for spatial dependence and propagate uncertainty in LM estimation into population-level inference. We develop a two-stage spatial Bayesian framework that incorporates a conditional autoregressive (CAR) prior to model dependence among neighbouring cortical vertices and a nested Markov chain Monte Carlo (MCMC) strategy to propagate uncertainty from subject-specific LM estimation into regression inference. A composite-basis representation enables computationally feasible population-level inference over the high-dimensional cortical surface. In simulation studies, the CAR prior substantially improved subject-level LM estimation accuracy, while CAR-based approaches reduced population-level estimation error and generally improved detection of localized group effects relative to independent estimation. Nested MCMC produced posterior mean estimates comparable to those of CAR plug-in, while yielding slightly wider credible intervals that account for uncertainty in subject-level LM estimation. Application to HCP-Aging resting-state fMRI data identified significant negative associations between age and LM in localized cortical regions, particularly in frontal and precentral areas. These findings demonstrate the benefits of spatial modeling for cortical LM estimation and uncertainty propagation for population-level Bayesian inference.
\end{abstract}

\section{Introduction}
\label{intro}

Functional magnetic resonance imaging (fMRI) provides a non-invasive measure of brain activity through changes in the blood-oxygen-level-dependent (BOLD) signal, which reflects hemodynamic responses associated with underlying neural activity (\citep{ogawa1990brain, logothetis2001neurophysiological}). Resting-state fMRI, in which participants lie quietly in the scanner without performing any explicit task, captures spontaneous fluctuations that reflect the brain's intrinsic functional organisation. These fluctuations exhibit long-memory and scale-free temporal structure, reflecting persistent and anti-persistent dynamics of ongoing neural activity \citep{he2011scale}. Characterizing this temporal dependence across cortical locations may provide insights into the spatial organization of intrinsic brain dynamics, while investigating its associations with age and other individual characteristics may help better understand population level heterogeneity in these dynamics.

Conventional short-range temporal dependence models, such as a first-order autoregressive process, may not adequately capture the temporal dependence observed in resting-state BOLD signals. In an AR(1) process, the autocovariance decays exponentially as \(\gamma(h)\propto\rho^{|h|}\), where \(h\) denotes the temporal lag and \(|\rho|<1\) controls the rate of decay \citep{woolrich2001temporal}. In contrast, long-memory processes exhibit slowly decaying temporal dependence, characterized by a power-law autocovariance \(\gamma(h)\propto h^{-\alpha_{\mathrm{cov}}}\) for large \(h\), where \(\alpha_{\mathrm{cov}}>0\) determines the rate of decay \citep{beran2017statistics, Knight_1997}. Wavelet-domain representations provide a natural framework for modelling this dependence across temporal scales \citep{bullmore2004wavelets}. Building on wavelet-based Bayesian approaches, \cite{Shahhosseini2025LM} proposed a two-stage framework that estimates voxelwise long-memory parameters and relates them to subject-level covariates through population-level regression. Applied to resting-state fMRI data from children and adolescents, this framework identified associations between age and long-memory structure, motivating further investigation of these associations in adult populations.

Spatial Bayesian models have been widely used in fMRI analysis to account for spatial dependence in brain activity. Conditional autoregressive (CAR), and related Markov random field priors have been used to regularize activation coefficients and autoregressive parameters in volumetric fMRI \citep{penny2005bayesian, quiros2010bayesian}, while surface-based Bayesian models have used sparse GMRF/SPDE priors to model task-activation fields across neighbouring cortical vertices \citep{mejia2020bayesian, spencer2022spatial}. Long-memory structure has also been incorporated into Bayesian fMRI models, including wavelet-domain regression with long-memory errors \citep{jeong2013wavelet} and a spatiotemporal model in which an MRF prior was placed on activation indicators while long-memory parameters were clustered through a Dirichlet process \citep{zhang2014spatio}. However, these approaches do not explicitly model spatial dependence in a vertexwise cortical LM field. The present work addresses this gap by placing a CAR prior directly on subject-specific LM maps estimated from resting-state fMRI, enabling spatial regularization at the cortical surface level while propagating Stage~1 posterior uncertainty into population-level inference through a nested MCMC strategy.

The nested MCMC strategy proposed here draws on the framework of cut-posterior inference in modular Bayesian models \citep{plummer2015cuts, frazier2025cutting}. Cut-posterior methods were developed for multi-stage models in which unrestricted feedback from a downstream module to an upstream module is undesirable, typically because misspecification in the second stage could affect first-stage parameter estimates. Under this formulation, the first-stage posterior is held fixed and the second stage is conditioned on draws from it, preserving a one-way flow of information between model components. Computational strategies for sampling from cut posteriors include nested MCMC schemes, in which retained draws from the first stage are passed into the second \citep{moss2024efficient}. This modular structure propagates first-stage posterior uncertainty into downstream inference without allowing the second-stage model to update first-stage parameter estimates. A further computational advantage is that second-stage sampling can be performed independently across retained first-stage posterior draws, allowing the computations to be distributed across multiple processors.

While the framework of \cite{Shahhosseini2025LM} provided the foundation for covariate-driven inference on long-memory structure in resting-state fMRI, two limitations remained. First, subject-level LM parameters were estimated without explicitly modelling spatial dependence between neighbouring cortical locations. Second, population-level analysis relied on a plug-in strategy that conditioned on posterior mean LM maps, treating these estimates as fixed and thereby failing to propagate Stage~1 posterior uncertainty into Stage~2 inference. Accounting for this uncertainty is particularly important when investigating spatially localized associations with age, as uncertainty in subject-level long-memory estimation may affect the precision of population-level estimates and the identification of cortical regions associated with individual characteristics. To address the first limitation, we introduce a conditional autoregressive (CAR) prior to model spatial dependence in the subject-level LM field. To address the second, we propose a nested MCMC strategy based on a modular cut-posterior formulation, allowing posterior uncertainty from Stage~1 to be propagated into population-level inference. This formulation preserves the one-way information flow of the original two-stage framework, so that subject-level LM estimates are not updated by the population-level regression model. Although a fully joint Bayesian formulation could propagate uncertainty across stages, it would require simultaneous sampling over the subject-level LM fields and population-level regression coefficients, a problem whose dimension grows with both the number of subjects and the number of cortical vertices and quickly becomes computationally prohibitive for the large surface datasets considered here. The proposed modular formulation avoids this by decoupling the two stages: Stage~1 is fitted once per subject and the retained posterior draws are passed forward, so that Stage~2 sampling operates only over the lower-dimensional population-level parameters. This preserves the scalability of the original two-stage framework while still propagating Stage~1 uncertainty into downstream inference. Together, these extensions provide spatially informed estimation and uncertainty propagation for covariate-driven inference on cortical long-memory structure.

We evaluate the proposed methodology through simulation studies and apply it to resting-state fMRI data from 567 adults aged 36–89 years in the Human Connectome Project Aging (HCP-Aging) to investigate associations between age and cortical long-memory structure. The analysis identifies spatially localized negative associations between age and the long-memory parameter, predominantly in frontal cortical regions, illustrating the ability of the proposed framework to characterize population-level associations in high-dimensional neuroimaging data while propagating uncertainty from subject-level estimation. The remainder of the paper is organized as follows. Section \ref{model} presents the statistical framework, including the subject-level spatial model, the population-level analysis, and the nested MCMC formulation. Section \ref{simulation_sec} evaluates the performance of the proposed methods through simulation studies. Section \ref{results_sec} applies the framework to the HCP-Aging data. Section \ref{discussion_sec} discusses the findings, methodological implications, limitations, and directions for future work.

\section{Statistical Modeling}\label{model}

We consider a two-stage Bayesian modeling strategy for estimating spatially varying long-memory (LM) parameters in resting-state fMRI. This framework separates subject-level estimation from population-level inference. In the first stage, subject-specific LM parameter maps are estimated using a spatial Bayesian model with a conditional autoregressive (CAR) prior. In the second stage, the estimated LM maps are related to subject-specific covariates through a group-level regression model. A nested MCMC strategy propagates posterior uncertainty from the first stage into the second.

\subsection{Long-memory Parametrization and Wavelet Representation}

To characterize the scale-free temporal dynamics of resting-state BOLD signals, we model the underlying time series at each cortical vertex as a long-range dependent (LRD) stochastic process. Formally, a stationary discrete-time process exhibits long-range dependence if its autocovariance function $\gamma(k)$ is non-summable, that is, $\sum_{k=-\infty}^{\infty} |\gamma(k)| = \infty$. A canonical parametric framework for capturing this property is fractional Gaussian noise (fGn), defined as the stationary increment process of fractional Brownian motion with Hurst parameter $H \in (0,1)$. For an fGn process with marginal variance $\sigma^2$, the autocovariance at temporal lag $k \in \mathbb{Z}$ is
\begin{equation}
\gamma(k) = \frac{\sigma^2}{2}\left( |k+1|^{2H} - 2|k|^{2H} + |k-1|^{2H} \right).
\label{eq:fgn_autocov}
\end{equation}
To obtain the asymptotic behavior as $k \to \infty$, we apply the binomial expansion to each shifted term. For large $k$,
\begin{equation*}
|k \pm 1|^{2H} = k^{2H}\!\left(1 \pm \frac{1}{k}\right)^{2H} = k^{2H}\left(1 \pm \frac{2H}{k} + \frac{2H(2H-1)}{2k^2} + O(k^{-3})\right).
\end{equation*}
Substituting into equation~\eqref{eq:fgn_autocov}, the leading-order terms cancel and the dominant contribution is
\begin{equation}
\gamma(k) \sim \sigma^2 H(2H-1)\, k^{2H-2} \quad \text{as } k \to \infty,
\label{eq:fgn_asymp}
\end{equation}
which is characteristic of long-memory processes when $H \in (1/2, 1)$ \citep{beran2017statistics, Knight_1997}. More broadly, semiparametric formulations characterize long-range dependence through the envelope $\gamma(k) \sim C k^{-\alpha_{\mathrm{cov}}}$ as $k \to \infty$, where $\alpha_{\mathrm{cov}} \in (0,1)$ is the covariance-decay exponent for the long-memory case. Comparing with equation~\eqref{eq:fgn_asymp} gives $\alpha_{\mathrm{cov}} = 2 - 2H$. This structure also admits a frequency-domain characterization: under standard regularity conditions, the spectral density, which is the Fourier transform of $\gamma$, satisfies $f(\omega) \sim C_f |\omega|^{-\alpha_{\mathrm{wav}}}$ as $\omega \to 0$, and the time- and frequency-domain decay exponents are related by $\alpha_{\mathrm{wav}} = 1 - \alpha_{\mathrm{cov}}$. Substituting $\alpha_{\mathrm{cov}} = 2 - 2H$ then yields $\alpha_{\mathrm{wav}} = 2H - 1$.

For high-dimensional cortical surface analysis, this scaling behaviour is most efficiently characterized in the wavelet domain. Let $\psi(t)$ denote a mother wavelet with $R$ vanishing moments satisfying $\int_{-\infty}^{\infty} t^r \psi(t)\, dt = 0$ for $r = 0, 1, \ldots, R-1$. When applied to a process whose spectral density follows $f(\omega) \sim C_f |\omega|^{-\alpha_{\mathrm{wav}}}$ as $\omega \to 0$, the wavelet coefficients $d_{j,k}$ at dyadic scale $j$ and translation $k$ satisfy, provided $R > (\alpha_{\mathrm{wav}} - 1)/2$,
\begin{equation}
\mathrm{Var}(d_{j,k}) \sim C_\psi\, 2^{j\alpha_{\mathrm{wav}}} \qquad \text{as } j \to \infty.
\end{equation}
Here, $C_\psi$ depends on the spectral intensity $C_f$ and the chosen wavelet basis \citep{flandrin1992wavelet, abry1998wavelet, Percival_Walden_2000}. For long-memory processes with $\alpha_{\mathrm{wav}} > 0$, the wavelet variance grows without bound as the scale $j$ increases, reflecting the concentration of energy at low frequencies that is characteristic of such processes. This scale-dependent growth is not a pathology but rather the key property that enables estimation: a log-linear regression of $\log \mathrm{Var}(d_{j,k})$ against $j$ has slope $\alpha_{\mathrm{wav}} \log 2$, providing a direct route to estimating the long-memory exponent from the empirical wavelet variances across scales \citep{abry1998wavelet}. The wavelet-domain scaling exponent satisfies
\begin{equation}
\alpha_{\mathrm{wav}} = 2H - 1 = 1 - \alpha_{\mathrm{cov}},
\end{equation}
which connects it directly to the Hurst parameter and the covariance-decay exponent.

Throughout the remainder of this paper, we let $\alpha$ denote $\alpha_{\mathrm{wav}} \in (-1,1)$, a parameterization that accommodates anti-persistent ($\alpha < 0$), white-noise ($\alpha = 0$), and long-memory ($\alpha > 0$) behaviour within a unified wavelet-domain framework, and our inferential target is the spatially varying field of these parameters across the cortical surface.

The wavelet-domain LM parameter is constrained to lie within the interval $(-1,1)$. Therefore, to facilitate estimation in an unconstrained space, we introduce the transformation
\begin{equation}
    \lambda_{s,v} = 2\tanh^{-1}(\alpha_{s,v}),
\end{equation}
with inverse mapping
\begin{equation}
    \alpha_{s,v} = \tanh\!\left(\frac{\lambda_{s,v}}{2}\right).
\end{equation}
Inference is therefore performed on the unconstrained parameter $\lambda_{s,v}$, while interpretation remains on the original LM scale through the inverse transformation.

Direct likelihood evaluation under long-memory covariance models is computationally demanding because long-memory dependence induces a dense covariance structure. To obtain a tractable representation, we apply the discrete wavelet transform (DWT) to each vertexwise time series \citep{mallat2002theory}. The DWT projects the signal onto a collection of multiscale basis functions indexed by temporal scale and location, yielding a transformed representation that separates signal variation across frequency bands.

Let $W$ denote the discrete wavelet transformation matrix associated with the chosen wavelet family and decomposition level. The rows of $W$ are constructed from discretized wavelet basis functions indexed by scale and temporal translation, together with the scaling functions at the coarsest resolution. These rows are organized across resolution levels, from finer scales representing higher-frequency variation to coarser scales representing lower-frequency variation. For a voxelwise time series $Y_{s,v}$ of length $T$, the wavelet-domain representation is

\begin{equation}
    Z_{s,v} = W Y_{s,v},
\end{equation}

\noindent where $W \in \mathbb{R}^{T^* \times T}$ and $Z_{s,v} \in \mathbb{R}^{T^*}$. Here, $T^*$ denotes the total number of wavelet coefficients produced by the transformation. Each row of $W$ corresponds to a wavelet or scaling coefficient associated with a particular resolution level and temporal location, while each column represents the contribution of an observation in the original time series to the transformed coefficients. The transformation is constructed from an orthogonal wavelet basis, while the dimensions of the resulting coefficient representation depend on the implementation and boundary treatment used in the DWT.

Under this representation, the covariance matrix transforms according to
\begin{equation}
    \Sigma_{s,v}^Z = W \Sigma_{s,v} W^\top.
\end{equation}
For long-memory processes, correlations between coefficients at distinct scales decay rapidly in the wavelet domain \citep{dijkerman1994correlation, fan2003approximate}. Consequently, the transformed covariance matrix becomes approximately diagonal with scale-dependent variance progression \citep{fadili2002wavelet, meyer2003wavelet, maxim2005fractional}. This property enables efficient likelihood evaluation while preserving the temporal dependence structure of the original signal.


\subsection{Stage 1: Subject-Level Spatial Model}\label{stage1}

Let $Y_{s,v} = (Y_{s,v,1}, \dots, Y_{s,v,T})^\top$ denote the BOLD time series of length $T$ observed at voxel $v$ for subject $s$. Following the wavelet-domain representation introduced above, let $Z_{s,v} = (Z_{s,v,1}, \dots, Z_{s,v,T^*})^\top$ denote the corresponding wavelet coefficients. Under the variance progression formulation of \cite{wornell2002wavelet}, the variance of the $i$-th wavelet coefficient is modeled as
\begin{equation}
    \mathrm{Var}(Z_{s,v,i}) = \nu_{s,v}\, 2^{-m_i \alpha_{s,v}},
\end{equation}
where $m_i$ denotes the associated wavelet scale, $\nu_{s,v} > 0$ is the innovation variance, and $\alpha_{s,v} \in (-1,1)$ denotes the wavelet-domain long-memory parameter at voxel $v$ for subject $s$, corresponding to the scale-dependent variance exponent defined above. Under the approximate decorrelation property of the DWT, the wavelet coefficients are modeled as conditionally independent Gaussian variables with zero mean and scale-dependent variances given above.

Spatial dependence among neighboring cortical vertices is incorporated through a conditional autoregressive (CAR) prior imposed on the transformed LM parameter field. Let \(\lambda_s=(\lambda_{s,1},\ldots,\lambda_{s,V})^\top\) collect the transformed LM parameters across cortical vertices for subject \(s\).

Following the neighbourhood-based formulation of \cite{besag1974spatial}, two vertices are defined as neighbours when they are directly connected by an edge of the cortical surface mesh. This neighbourhood definition is particularly well suited to the present data, as the HCP preprocessing pipeline maps cortical fMRI signals from the grey-matter ribbon onto reconstructed cortical surfaces and resamples them to a common standard surface mesh with vertexwise correspondence across subjects \citep{glasser2013minimal}. 

The neighbourhood structure is encoded by the symmetric adjacency matrix $A$, with $A_{vu} = 1$ if vertices $v$ and $u$ are connected by a mesh edge and $A_{vu} = 0$ otherwise, with $A_{vv} = 0$. The corresponding degree matrix $D$ is diagonal, with
\begin{equation}
    D_{vv} = \sum_{u=1}^{V} A_{vu},
\end{equation}
so that $D_{vv}$ gives the number of immediate neighbours of vertex $v$. The CAR precision matrix for subject $s$ is then defined as
\begin{equation}
    Q(\kappa_s) = D - \kappa_s A,
\end{equation}
and the spatial prior is
\begin{equation}
    \boldsymbol{\lambda}_s \mid \sigma_s^2, \kappa_s \sim \mathcal{N}\!\left(0,\, \sigma_s^2 Q(\kappa_s)^{-1}\right).
\end{equation}
Here, $\sigma_s^2$ controls the overall spatial variability, while 
$\kappa_s \in [0,1)$ controls the strength of dependence between 
neighbouring vertices, with $\kappa_s = 0$ reducing the prior to 
one with no spatial dependence between vertices; in practice we 
assign $\kappa_s$ a $\mathrm{Beta}(a,b)$ prior, which assigns zero 
mass to the boundary.

Prior distributions are specified for both vertexwise variance components and spatial hyperparameters to complete the Bayesian hierarchy. For each subject $s$ and vertex $v$, the innovation variance is assigned an independent inverse-gamma prior \[\nu_{s,v} \sim \text{Inv-Gamma}(a_\nu, b_\nu).\] 

The subject-specific spatial variance and dependence parameters are assigned the priors
\[
\begin{aligned}
\sigma_s^2&\sim\operatorname{Inv\text{-}Gamma}(a_\sigma,b_\sigma),\\
\kappa_s&\sim\operatorname{Beta}(a_\kappa,b_\kappa).
\end{aligned}
\]

The quantities \(a_\nu,b_\nu,a_\sigma,b_\sigma,a_\kappa,\) and \(b_\kappa\) denote fixed hyperparameters controlling prior concentration and scale. Weakly informative prior specifications are adopted for the spatial hyperparameters to avoid overly restrictive smoothing assumptions. Specifically, we set \((a_\sigma,b_\sigma)=(2,1)\) and \((a_\kappa,b_\kappa)=(1,1)\). For the voxelwise innovation variance, we use \((a_\nu,b_\nu)=(2,2)\), consistent with the prior specification adopted in \citep{Shahhosseini2025LM}.

\subsection{ Stage 2: Group-Level Regression Analysis}\label{stage2}

Following subject-level estimation, the second stage relates the long-memory parameter maps to subject-specific covariates through a reduced-dimensional spatial regression framework. Rather than conditioning on a single posterior summary from Stage~1, the nested MCMC formulation propagates retained posterior realizations of the subject-specific LM fields into the group-level analysis. 

For subject $s = 1, \ldots, S$, let
\begin{equation}
    \boldsymbol{\alpha}_s^{(t)} = \left(\alpha_{s,1}^{(t)}, \ldots, \alpha_{s,V}^{(t)}\right)^\top \in \mathbb{R}^{V}
\end{equation}
denote the sampled LM parameter map corresponding to retained Stage~1 iteration $t$. The subject-specific maps are then assembled as
\begin{equation} \label{eq14}
    \boldsymbol{\alpha}^{(t)} = \begin{pmatrix} \boldsymbol{\alpha}_1^{(t)\top} \\ \vdots \\ \boldsymbol{\alpha}_S^{(t)\top} \end{pmatrix} \in \mathbb{R}^{S \times V}.
\end{equation}

At each nested iteration, the subject-specific LM realizations are collected across all subjects to form an $S \times V$ matrix, with subjects indexed by rows and spatial locations by columns. This matrix serves as the response in the Stage 2 regression model.

Direct regression in vertex space is computationally challenging because of the high dimensionality of the cortical surface. To obtain a tractable population-level representation while retaining the dominant spatial structure of the subject-level LM maps, we employ the composite-basis dimension-reduction framework of \cite{miranda2023novel}. This framework was also used in \cite{Shahhosseini2025LM}, but is now applied to posterior realizations of the subject-specific LM fields rather than to a single estimated map. At each retained Stage 1 iteration, the composite basis is constructed from the corresponding realization of the subject-specific LM maps, allowing the dimension-reduction step to reflect variability in the estimated LM fields. The role of this basis is distinct from that of the Stage 1 CAR prior: the CAR prior spatially regularizes subject-specific LM estimation through local cortical-mesh dependence, whereas the composite basis provides a lower-dimensional representation for population-level regression.

The brain volume is first partitioned into anatomical regions of interest using the Schaefer atlas \citep{schaefer2018local}. Within each region $g$, singular value decomposition (SVD) is applied to extract orthonormal local basis functions describing dominant within-region variability. Let $\Phi_g^{(t)}$ denote the retained iteration-specific basis matrix for region $g$, where components explaining at least 99\% of the variance are preserved. These regional bases are assembled into a block-diagonal matrix
\begin{equation}
    \Phi^{(t)}= \text{blkdiag}(\Phi_1^{(t)}, \dots, \Phi_G^{(t)}),
\end{equation}
where $G$ denotes the number of anatomical regions.

A second-level SVD is then applied to the projected data to obtain a global basis $\Psi^{(t)}$, capturing dependence across regions. Combining these two stages yields the composite basis
\begin{equation}
    \Upsilon{(t)} = \Phi^{(t)}\Psi^{(t)},
\end{equation}
which jointly represents local anatomical structure and global spatial variation. Projecting the LM maps onto this reduced basis produces
\begin{equation}
    \boldsymbol{\alpha}^{*(t)} = \boldsymbol{\alpha}^{(t)}\Upsilon^{(t)},
\end{equation}
where $\alpha^{*(t)} \in \mathbb{R}^{S \times PC_f^{(t)}}$ denotes the reduced-dimensional representation of the subject-specific LM maps at nested iteration $t$, and $PC_f^{(t)}$ is the number of retained composite components.

Group-level inference is performed through the regression model

\begin{equation}
    \boldsymbol{\alpha}^{*\,(t)} = X\boldsymbol{\beta}^{*\,(t)} + \boldsymbol{\varepsilon}^{*\,(t)}
\end{equation}

where $X$ is the $S \times P$ design matrix of subject-level 
covariates, $\boldsymbol{\beta}^{*\,(t)}$ is a $P \times PC_f^{(t)}$ 
coefficient matrix, and $\varepsilon_r^{*\,(t)}$ is a random  error vector  for basis component $r$ at Stage~1 iteration $t$. 
For each retained component $r$,
\begin{equation}
    \varepsilon_r^{*\,(t)} \overset{\text{ind}}{\sim} 
    \mathcal{N}\!\left(0,\, \delta_r^{2\,(t)} I\right), 
    \qquad r = 1, \dots, PC_f^{(t)}.
\end{equation}
Prior distributions are assigned independently across components as
\begin{equation}
    \beta_r^* \mid \delta_r^2 
    \overset{\text{ind}}{\sim} \mathcal{N}(\mu_0,\, \delta_r^2 \Lambda_0),
\end{equation}
\begin{equation}
    \delta_r^2 
    \overset{\text{ind}}{\sim} \operatorname{Inv-Gamma}(k, l).
\end{equation}

Conditional on each retained Stage 1 realization, posterior inference proceeds via Gibbs sampling in the reduced basis space. Posterior samples of the regression coefficients are projected back to vertex space through the composite basis transformation, yielding spatially varying coefficient maps. Detailed derivations of the posterior updates are provided in \ref{appendixA.1}. The nested MCMC procedure used to propagate uncertainty across the two stages is described in the next section.

\subsection{Nested MCMC} \label{nested}

Inference for the two-stage CAR framework is performed using a 
modular Bayesian strategy with a cut-posterior structure, in which 
information is propagated from Stage~1 to Stage~2 without feedback 
from the population-level model to the subject-level estimation 
\citep{plummer2015cuts, frazier2025cutting}. Cut-posterior methods 
provide a general framework for restricting feedback between 
components of a modular Bayesian model while retaining conditional 
dependence between successive modules. In the present setting, the 
motivation is to propagate subject-level posterior variability 
through the two-stage analysis.

A conventional plug-in approach conditions the population-level 
model on a fixed posterior summary of the subject-specific LM field, 
such as $E(\boldsymbol{\alpha}_s \mid \mathbf{Z}_s)$, and therefore 
does not carry the full Stage~1 posterior distribution into Stage~2. 
Instead, we use a nested MCMC procedure in which retained posterior 
realizations of the subject-level LM field are successively 
propagated into the group-level regression model, following the 
general computational structure used for cut-posterior inference 
\citep{moss2024efficient}. This formulation preserves the one-way 
modular structure of the two-stage analysis while allowing 
variability in the subject-level LM estimates to contribute directly 
to population-level inference.

Let $\mathbf{Z} = \{\mathbf{Z}_s\}_{s=1}^S$ denote the collection 
of subject-level wavelet-domain data, and let 
$\boldsymbol{\alpha} = \{\boldsymbol{\alpha}_s\}_{s=1}^S$ denote the 
corresponding subject-specific LM fields, where 
$\boldsymbol{\alpha}_s = (\alpha_{s,1}, \ldots, \alpha_{s,V})^\top$. 
Let $X$ denote the population-level design matrix and let 
$\boldsymbol{\beta}$ collect the corresponding regression parameters. 
Under the full joint posterior, Stage~1 would condition on both 
$\mathbf{Z}$ and the population-level parameters $\boldsymbol{\beta}$; 
the cut-posterior severs this feedback by replacing the full 
conditional $\pi(\boldsymbol{\alpha} \mid \mathbf{Z}, X, \boldsymbol{\beta})$ 
with $\pi_1(\boldsymbol{\alpha} \mid \mathbf{Z})$, which depends only 
on the subject-level data. The inferential target is therefore the 
cut-posterior factorization
\begin{equation}
    \pi_{\mathrm{cut}}(\boldsymbol{\alpha}, \boldsymbol{\beta} 
    \mid \mathbf{Z}, X) 
    = \pi_1(\boldsymbol{\alpha} \mid \mathbf{Z})\, 
      \pi_2(\boldsymbol{\beta} \mid \boldsymbol{\alpha}, X),
\end{equation}
where $\pi_1(\boldsymbol{\alpha} \mid \mathbf{Z})$ is the Stage~1 
posterior arising from the subject-level model alone,
\begin{equation}
    \pi_1(\boldsymbol{\alpha} \mid \mathbf{Z}) \propto 
    \prod_{s=1}^{S} \pi(\mathbf{Z}_s \mid \boldsymbol{\alpha}_s)\, 
    \pi(\boldsymbol{\alpha}_s),
\end{equation}
which does not depend on $\boldsymbol{\beta}$ or $X$, so that 
information flows from Stage~1 to Stage~2 but not in the reverse 
direction. This construction follows the general cut-posterior 
framework \citep{plummer2015cuts, frazier2025cutting}, though here 
the cut structure is used to preserve the modular two-stage analysis 
and propagate Stage~1 uncertainty, rather than to protect the first 
module from anticipated misspecification in the second.

For each retained Stage~1 iteration $t = 1, \ldots, M$, the 
retained draw $\boldsymbol{\alpha}^{(t)}$ is assembled from the 
subject-specific LM realizations as defined in 
Equation~\eqref{eq14}. The corresponding composite basis 
$\boldsymbol{\Upsilon}^{(t)}$ is constructed and used to obtain the 
reduced-dimensional representation 
$\boldsymbol{\alpha}^{*(t)} = \boldsymbol{\alpha}^{(t)} 
\boldsymbol{\Upsilon}^{(t)}$. Conditional on 
$\boldsymbol{\alpha}^{*(t)}$, the Stage~2 regression model is fitted 
using the population-level design matrix $X$, producing posterior 
draws of the corresponding regression parameters. Repeating this 
procedure across retained Stage~1 realizations yields a nested MCMC 
scheme in which each Stage~2 analysis is conditional on a different 
draw from the Stage~1 posterior, thereby propagating the 
distribution of the subject-level LM fields through the 
population-level model rather than conditioning on a single fixed 
summary \citep{moss2024efficient}.

After projecting the conditional Stage~2 posterior draws back to 
vertex space using their corresponding composite bases, the resulting 
posterior distributions provide a Monte Carlo approximation to the 
marginal cut posterior, justified as follows. Specifically, if 
$\boldsymbol{\alpha}^{(t)}$, $t = 1, \ldots, M$, denotes the 
collection of draws retained from 
$\pi_1(\boldsymbol{\alpha} \mid \mathbf{Z})$ across Stage~1, then 
by standard Monte Carlo integration,
\begin{equation}
    \pi_{\mathrm{cut}}(\boldsymbol{\beta} \mid \mathbf{Z}, X) 
    = \int \pi_2(\boldsymbol{\beta} \mid \boldsymbol{\alpha}, X)\, 
      \pi_1(\boldsymbol{\alpha} \mid \mathbf{Z})\, d\boldsymbol{\alpha} 
    \approx \frac{1}{M}\sum_{t=1}^{M} 
      \pi_2\!\left(\boldsymbol{\beta} \mid 
      \boldsymbol{\alpha}^{(t)}, X\right),
\end{equation}
where each term 
$\pi_2(\boldsymbol{\beta} \mid \boldsymbol{\alpha}^{(t)}, X)$ is 
itself a posterior distribution over $\boldsymbol{\beta}$ obtained 
by running Stage~2 Gibbs sampling conditional on the fixed 
realization $\boldsymbol{\alpha}^{(t)}$.

Thus, population-level inference is based on the mixture of 
conditional Stage~2 posteriors obtained across Stage~1 realizations, 
rather than on a regression conditioned on a single estimated LM 
map. The resulting vertexwise posterior distributions characterize 
the covariate effects while incorporating variability associated 
with subject-level LM estimation. In this way, the nested MCMC 
procedure preserves the modular separation between subject- and 
population-level models while carrying posterior uncertainty from 
Stage~1 through to the final spatial inference 
\citep{plummer2015cuts, frazier2025cutting, moss2024efficient}.

\section{Simulation Study}
\label{simulation_sec}

The simulation was designed to address three complementary questions. First, we assessed whether the model could accurately recover subject-specific long-memory fields from the simulated wavelet coefficients. Second, we examined whether incorporating spatial dependence through the CAR prior and propagating Stage~1 uncertainty through the nested MCMC strategy affected population-level estimation, uncertainty quantification, and detection of a localized group effect. Third, we evaluated the robustness of the three analysis strategies under increasing observation noise. These comparisons allow us to assess the contribution of spatial regularization under plug-in inference and the additional contribution of uncertainty propagation under the CAR model.

To reflect the spatial organization of the real brain data, we constructed the simulations on a two-dimensional gray-matter mask containing 1,268 spatial locations. We mapped a cortical parcellation atlas onto the same grid, allowing the simulated locations to retain anatomical parcel labels and ensuring compatibility with the parcel-based dimension-reduction procedure used in the subsequent two-stage analysis. We introduced subject-specific variation in the LM field through a spatially structured latent process, producing locally smooth patterns. Each simulated dataset consisted of 200 subjects, balanced as 100 subjects in each of two groups. Three scenarios were considered: a null scenario with no added group effect, a weak-signal scenario with signal strength $b_0 = 1.2$, and a strong-signal scenario with $b_0 = 1.6$, each with 100 independent replications.

For each subject, we generated the latent spatial field $\boldsymbol{\lambda}_s \sim \text{CAR}(\kappa = 0.85)$ using a conditional autoregressive model with four-neighbor adjacency and transformed it to the LM parameter via $\alpha_{s,v} = \tanh(\lambda_{s,v}/2)$, restricting $\alpha$ to the interval $(-1,1)$. Wavelet-domain observations were generated from the resulting subject-specific LM maps according to the same variance progression model used in the likelihood, with $\nu = 1$. For subjects in the signal group, a localized increment was added on the $\lambda$-scale before transformation:
\begin{equation}
    b_v = b_0 \exp\!\left(-\frac{d_v^2}{2\sigma_{\text{graph}}^2}\right),
\end{equation}
where $d_v$ denotes the graph distance from vertex $v$ to the center of the active region and $\sigma_{\text{graph}} = 2.0$ controls the spatial decay, so that the signal-group LM field becomes $\alpha_{s,v} = \tanh\!\bigl((\lambda_{s,v} + b_v)/2\bigr)$. The parameter $b_0 \in \{1.2, 1.6\}$ sets the overall signal strength. To assess robustness under observation noise, independent Gaussian noise $\varepsilon_{s,v,i} \sim \mathcal{N}(0, C^2)$ with $C \in \{0, 0.5, 1.0\}$ was added to the wavelet coefficients prior to Stage~1 estimation.

\subsection{Subject-level recovery}

We first evaluated the ability of the independent and CAR models to recover subject-specific LM fields. We assessed estimation accuracy using root mean squared error (RMSE), computed across all 1,268 spatial locations (overall RMSE) and within the 30 locations with a nonzero group effect (active-region RMSE). We computed both measures for each subject and averaged them across subjects and the 100 simulation replications. We compared the models across signal strengths $b_0 \in \{1.2, 1.6\}$ and additive noise levels $C \in \{0, 0.5, 1.0\}$.

\begin{figure*}[!htbp]
\centerline{\includegraphics[width=0.9\textwidth]{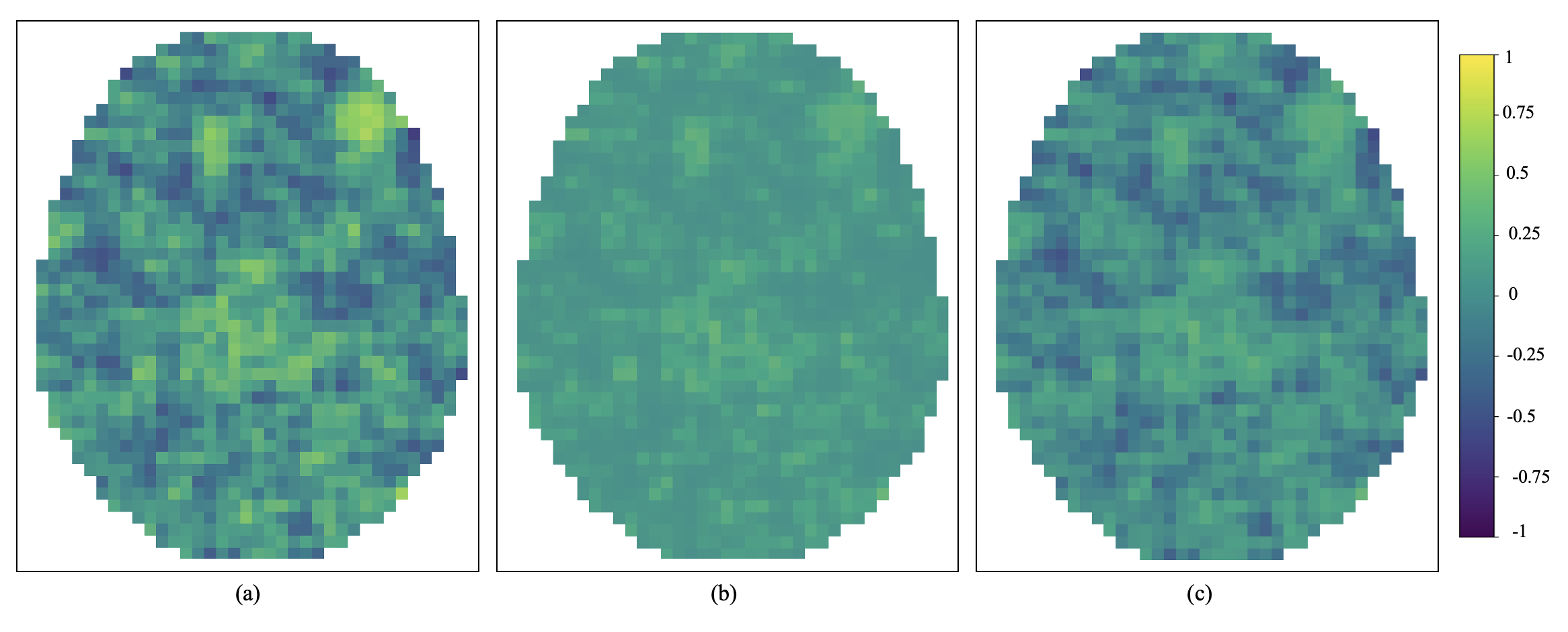}}
\caption{Spatial long-memory (LM) parameter maps for a selected simulated subject under signal strength $b_0 = 1.6$ and noise level $C = 0.5$. Panel~(a) shows the true LM parameter map. Panel~(b) shows the posterior mean LM estimates obtained using the independent model. Panel~(c) shows the posterior mean LM estimates obtained using the CAR model.\label{twostage_sim}}
\end{figure*}

Table~\ref{tab:rmse_comparison} summarizes the results. At $C=0$, the CAR model substantially reduced RMSE relative to the independent model across both signal strengths, with overall RMSE decreasing from approximately 0.204 to 0.054 and active-region RMSE from approximately 0.166 to 0.053. As $C$ increased, RMSE increased under both models, but the CAR model maintained lower RMSE across all conditions. At $C = 1.0$, overall RMSE was approximately 30\% lower under the CAR model, although the difference in active-region RMSE was smaller. Figure~\ref{twostage_sim} illustrates subject-level LM recovery for a representative simulated subject.

\begin{table}[htbp]
\centering
\begin{tabular}{llcccc}
\toprule
& & \multicolumn{2}{c}{Overall RMSE}
    & \multicolumn{2}{c}{Active-region RMSE} \\
\cmidrule(lr){3-4} \cmidrule(lr){5-6}
\textbf{$b_0$} & \textbf{Noise $C$} & \textbf{Independent} & \textbf{CAR} & \textbf{Independent} & \textbf{CAR} \\
\midrule
1.2 & 0   & 0.2043 & 0.0538 & 0.1657 & 0.0534 \\
    & 0.5 & 0.2200 & 0.1063 & 0.2157 & 0.1548 \\
    & 1.0 & 0.2469 & 0.1734 & 0.2746 & 0.2435 \\
\midrule
1.6 & 0   & 0.2045 & 0.0537 & 0.1666 & 0.0528 \\
    & 0.5 & 0.2205 & 0.1070 & 0.2304 & 0.1744 \\
    & 1.0 & 0.2477 & 0.1744 & 0.2979 & 0.2700 \\
\bottomrule
\end{tabular}
\caption{Subject-level RMSE of the long-memory parameter under
independent and CAR models, across signal strengths $b_0$ and
noise levels $C$. Overall RMSE is computed across all 1,268
spatial locations, while active-region RMSE is restricted to
the 30 locations with a nonzero group effect.}
\label{tab:rmse_comparison}
\end{table}

\subsection{Group-Effect Recovery}

Next, we evaluated population-level group-effect recovery across the three analysis strategies. We assessed estimation accuracy using the root mean squared error (RMSE) of the estimated group-effect coefficient $\beta$, computed across all spatial locations (overall RMSE) and within the 30 locations with a nonzero group effect (active-region RMSE). We assessed uncertainty quantification using the empirical coverage and mean width of the pointwise 95\% posterior credible intervals. Finally, we evaluated spatial detection using the proportion of locations with a nonzero group effect identified as significant after SimBaS correction at \(\varsigma^*<0.05\), along with the number of false positives.

\begin{figure*}[h]
\centerline{\includegraphics[width=0.9\textwidth]{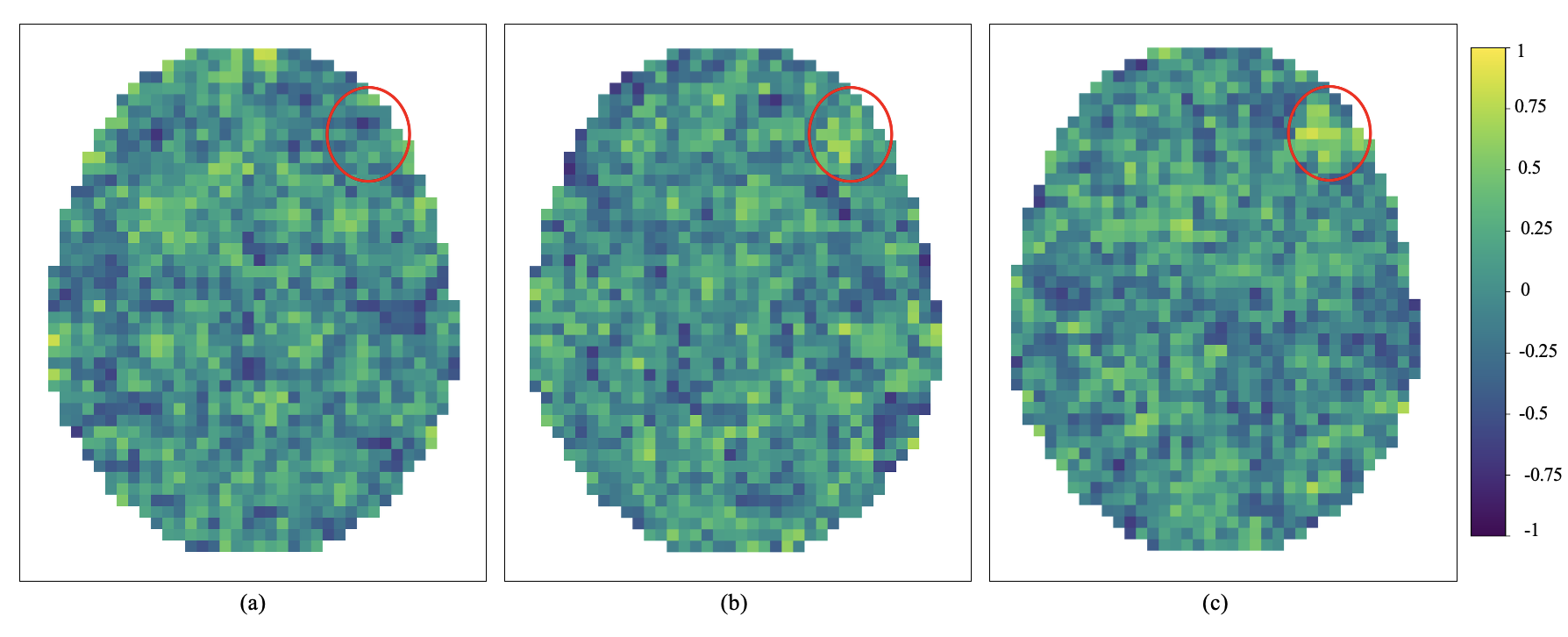}}
\caption{Simulated long-memory maps with localized spatial group effects. (a) Baseline LM map with no group effect; the red circle identifies the 30 connected locations where the group effect is introduced. (b) Group-averaged LM map for $b_0 = 1.2$. (c) Group-averaged LM map for $b_0 = 1.6$. Both non-null scenarios are shown with no additional noise ($C=0$).}\label{LM_sim}
\end{figure*}

We introduced the group effect within 30 connected spatial locations, with its magnitude decreasing smoothly with graph distance from the center. Figure~\ref{LM_sim}(a) identifies this region, while Figures~\ref{LM_sim}(b) and (c) show the resulting group-averaged LM maps for $b_0=1.2$ and $b_0=1.6$, respectively, with no additional noise $C=0$. The parameter $b_0$ controls the overall signal strength on the latent $\lambda$ scale.

Population-level estimation results are summarized in Table~\ref{tab:beta_recovery}. Under correct model specification ($C=0$), both CAR-based approaches substantially reduced overall and active-region $\beta$ RMSE compared with the independent plug-in model. For $b_0=1.2$, overall RMSE decreased from 0.0274 under independent plug-in to 0.0089 under CAR plug-in and 0.0082 under CAR nested, with similar improvements for $b_0=1.6$. CAR plug-in and CAR nested produced nearly identical posterior mean estimates, with nested MCMC yielding slightly wider credible intervals. Both CAR-based approaches achieved 100\% empirical pointwise coverage under correct model specification. When additional noise was introduced ($C=0.5$ and $C=1$), resulting in model misspecification, estimation error increased for all methods, but the CAR-based approaches maintained lower RMSE than independent plug-in.
 Mean credible interval widths decreased as $C$ increased, even though RMSE increased, suggesting sensitivity of posterior uncertainty estimates to model misspecification.

\begin{figure*}[h]
\centerline{\includegraphics[width=0.9\textwidth]{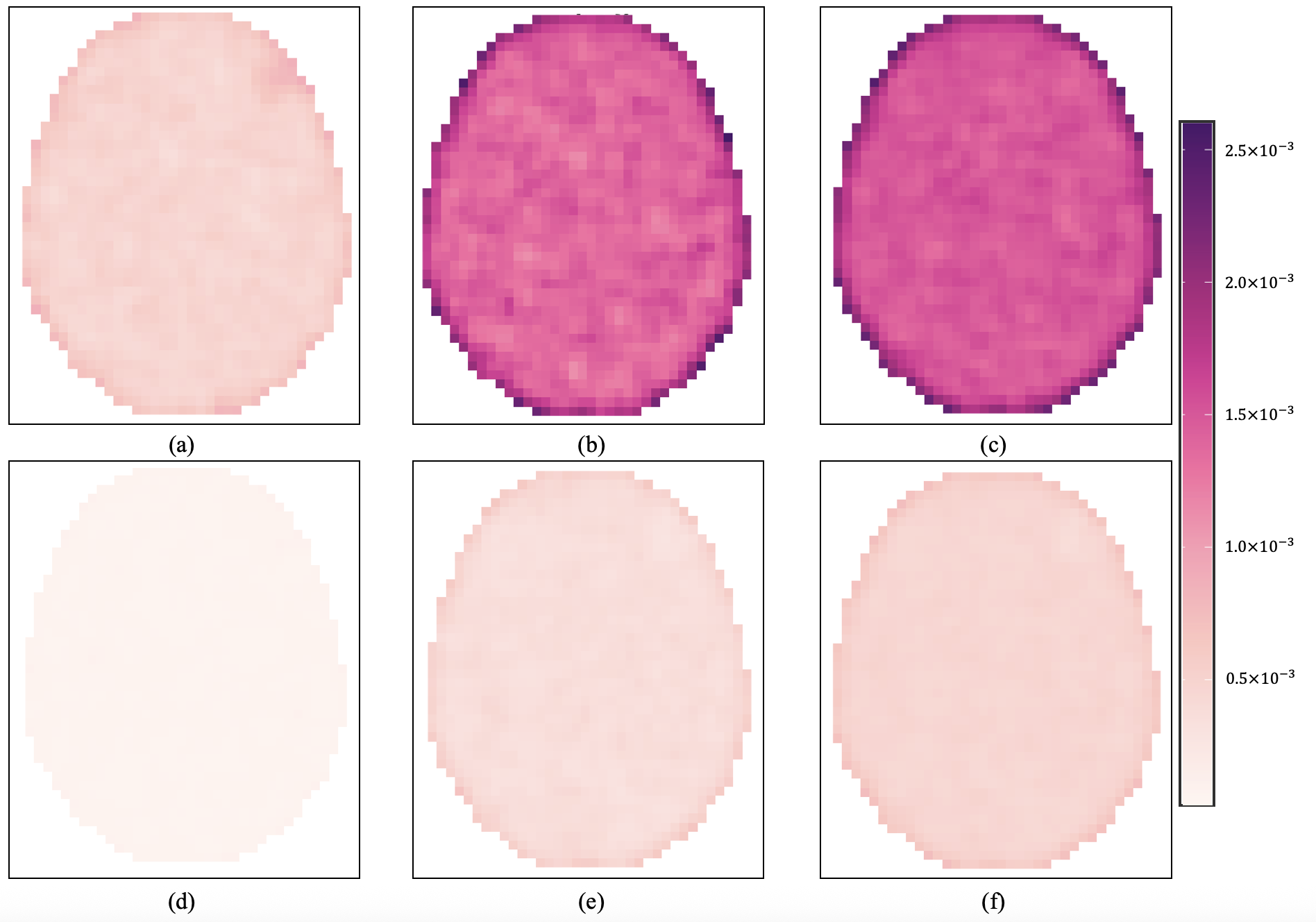}}
\caption{Simulation results. Posterior variance of the group-effect coefficient $\beta$ at signal strength $b_0=1.6$, under correct model specification ($C=0$, top row) and additional noise introducing model misspecification ($C=1$, bottom row), for independent plug-in (a, d), CAR plug-in (b, e), and CAR nested (c, f).}
\label{var_maps}
\end{figure*}

 Figure~\ref{var_maps} illustrates the spatial distribution of the posterior variance of the group-effect coefficient $\beta$ at $b_0=1.6$. Under correct model specification ($C=0$), both CAR-based approaches produced higher posterior variances than independent plug-in, with CAR nested yielding slightly higher variance than CAR plug-in. When additional noise introduced model misspecification ($C=1$), posterior variances decreased across all three approaches.

\begin{table}[htbp]
    \centering
    \resizebox{\textwidth}{!}{%
    \begin{tabular}{llcccccccccccc}
        \toprule
        & & \multicolumn{4}{c}{\textbf{Independent plug-in}} & \multicolumn{4}{c}{\textbf{CAR plug-in}} & \multicolumn{4}{c}{\textbf{CAR nested}} \\
        \cmidrule(lr){3-6} \cmidrule(lr){7-10} \cmidrule(lr){11-14}
        \textbf{$b_0$} & \textbf{$C$} & \textbf{$\beta$ RMSE} & \textbf{Active RMSE} & \textbf{Cov.\ (\%)} & \textbf{CI width} & \textbf{$\beta$ RMSE} & \textbf{Active RMSE} & \textbf{Cov.\ (\%)} & \textbf{CI width} & \textbf{$\beta$ RMSE} & \textbf{Active RMSE} & \textbf{Cov.\ (\%)} & \textbf{CI width} \\
        \midrule
        1.2 & 0   & 0.0274 & 0.0871 & 89.5  & 0.0854 & 0.0089 & 0.0132 & 100.0 & 0.1450 & 0.0082 & 0.0126 & 100.0 & 0.1546 \\
            & 0.5 & 0.0412 & 0.1862 & 60.4  & 0.0524 & 0.0232 & 0.1155 & 98.0  & 0.1098 & 0.0229 & 0.1150 & 98.3  & 0.1199 \\
            & 1.0 & 0.0511 & 0.2429 & 31.9  & 0.0310 & 0.0374 & 0.1874 & 85.3  & 0.0739 & 0.0375 & 0.1882 & 91.4  & 0.0857 \\
        \midrule
        1.6 & 0   & 0.0288 & 0.1081 & 91.2  & 0.0857 & 0.0090 & 0.0139 & 100.0 & 0.1458 & 0.0082 & 0.0139 & 100.0 & 0.1546 \\
            & 0.5 & 0.0482 & 0.2474 & 61.0  & 0.0527 & 0.0290 & 0.1618 & 98.0  & 0.1107 & 0.0288 & 0.1612 & 98.2  & 0.1198 \\
            & 1.0 & 0.0605 & 0.3215 & 31.7  & 0.0312 & 0.0460 & 0.2553 & 86.3  & 0.0748 & 0.0459 & 0.2548 & 91.1  & 0.0857 \\
        \bottomrule
    \end{tabular}}
    \caption{Population-level estimation and uncertainty quantification for the group-effect coefficient $\beta$ across the independent plug-in, CAR plug-in, and CAR nested approaches, signal strengths $b_0 \in \{1.2, 1.6\}$, and additive noise levels $C \in \{0, 0.5, 1.0\}$. $\beta$ RMSE reports the root mean squared error of the posterior mean across all 1,268 spatial locations; Active RMSE reports the same quantity restricted to the 30 true active vertices. Cov.\ (\%) denotes the observed pointwise 95\% posterior credible interval coverage, and CI width reports the mean width of those intervals across all spatial locations. All measures are averaged across 100 independent simulation replications.}
    \label{tab:beta_recovery}
\end{table}

\begin{figure*}[h]
\centerline{\includegraphics[width=0.9\textwidth]{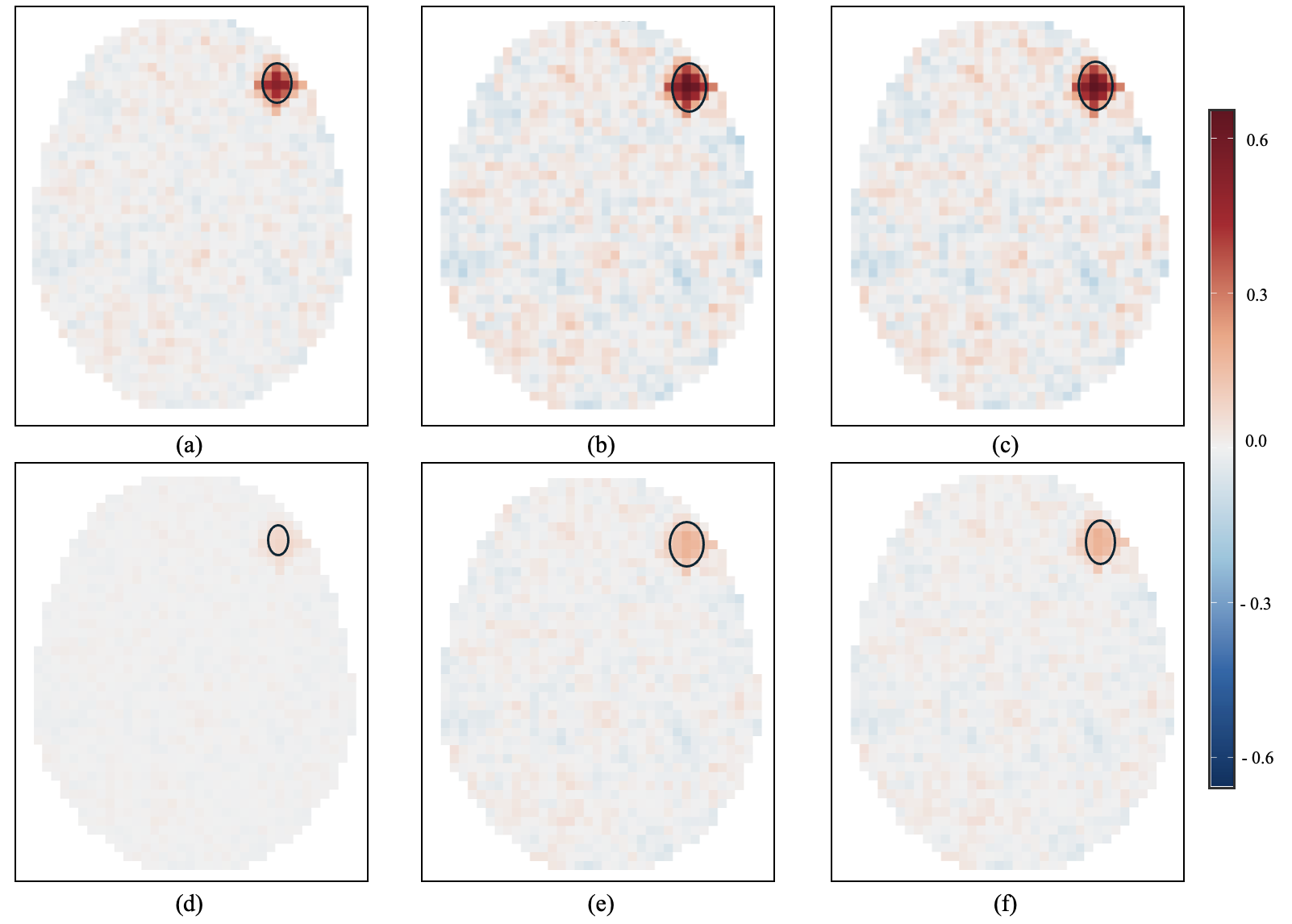}}
\caption{Simulation results. Posterior mean of the group-difference coefficient $\widehat{\beta}_v$ at signal strength $b_0 = 1.6$ under clean observations ($C = 0$, top row) and high observation noise ($C = 1.0$, bottom row), for the independent plug-in (a, d), CAR plug-in (b, e), and CAR nested (c, f) approaches. Black circles indicate vertices declared significant after SimBaS correction at $\varsigma^* < 0.05$.\label{results}}
\end{figure*}

We next examined spatial detection following SimBaS correction at $\varsigma^*<0.05$ (Table~\ref{tab:detection} and Figure~\ref{results}). Unlike estimation accuracy, spatial detection also depends on posterior uncertainty, making it important to distinguish detection sensitivity from uncertainty propagation. Under correct model specification ($C=0$), CAR plug-in and CAR nested recovered 73.3\% of the locations with a nonzero group effect at $b_0=1.2$, compared with 66.7\% for independent plug-in. At $b_0=1.6$, recovery rates were 80.0\%, 76.7\%, and 76.7\% for CAR plug-in, CAR nested, and independent plug-in, respectively. Under model misspecification ($C=0.5$ and $C=1$), CAR plug-in generally achieved the highest recovery rates. False-positive detections were rare across all conditions. These results indicate that propagating Stage~1 uncertainty does not necessarily increase detection sensitivity, even when posterior mean estimation accuracy remains comparable.

\begin{table}[htbp]
    \centering
    \begin{tabular}{llcccccc}
        \toprule
        & & \multicolumn{2}{c}{\textbf{Independent plug-in}} & \multicolumn{2}{c}{\textbf{CAR plug-in}} & \multicolumn{2}{c}{\textbf{CAR nested}} \\
        \cmidrule(lr){3-4} \cmidrule(lr){5-6} \cmidrule(lr){7-8}
        \textbf{$b_0$} & \textbf{$C$} & \textbf{Recovery (\%)} & \textbf{FP} & \textbf{Recovery (\%)} & \textbf{FP} & \textbf{Recovery (\%)} & \textbf{FP} \\
        \midrule
        1.2 & 0   & 66.7 & 0 & 73.3 & 0 & 73.3 & 0 \\
            & 0.5 & 63.3 & 0 & 70.0 & 1 & 60.0 & 0 \\
            & 1.0 & 46.7 & 0 & 63.3 & 0 & 46.7 & 0 \\
        \midrule
        1.6 & 0   & 76.7 & 0 & 80.0 & 0 & 76.7 & 0 \\
            & 0.5 & 73.3 & 0 & 76.7 & 0 & 73.3 & 0 \\
            & 1.0 & 56.7 & 0 & 73.3 & 0 & 63.3 & 0 \\
        \bottomrule
    \end{tabular}
    \caption{Stage~2 spatial detection performance for the independent plug-in, CAR plug-in, and CAR nested approaches across signal strengths $b_0 \in \{1.2, 1.6\}$ and additive noise levels $C \in \{0, 0.5, 1.0\}$. Recovery rate reports the percentage of the 30 true active spatial locations correctly identified as significant after SimBaS correction at $\varsigma^* < 0.05$. FP denotes the mean number of false-positive detections outside the true active region. All measures are averaged across 100 independent simulation replications.}
    \label{tab:detection}
\end{table}

Together, the simulation results demonstrate the benefits of the CAR prior for subject-level LM estimation and population-level group-effect recovery. The CAR-based approaches consistently achieved lower RMSE than the independent plug-in model, including under additional noise that introduced model misspecification. Nested MCMC propagated Stage~1 posterior uncertainty into population inference, producing wider credible intervals with little change in posterior mean accuracy. These findings motivate the application of the proposed framework to resting-state fMRI data.

\section{Real Data Application}\label{results_sec}
\subsection{HCP-A Data and Preprocessing}
We analyze data from the Human Connectome Project Aging (HCP-A), publicly available through the NIMH Data Archive and described in detail in \cite{bookheimer2019lifespan}. The HCP-A dataset consists of adults aged 36 to 100 years, with the goal of characterizing brain structure and function across the adult lifespan. Resting-state functional MRI (rs-fMRI) data were collected alongside an extensive battery of cognitive, behavioral, and health measures.

Resting-state fMRI data were acquired on a 3T Siemens Prisma scanner using a multiband gradient-echo EPI sequence and processed using the HCP minimal preprocessing pipeline \citep{glasser2013minimal}, which includes gradient distortion correction, motion correction, and registration to the MNI152 template. The HCP-A resting-state acquisition follows the Lifespan HCP protocol \citep{harms2018extending}, with a repetition time of TR $= 0.8$ s, echo time TE $= 37$ ms, and 2 mm isotropic spatial resolution. Rather than analysing signals in volumetric space, where normalisation and smoothing can mix signals across anatomical boundaries and opposing sulcal banks \citep{coalson2018impact}, cortical surface data were projected onto the fsLR 32k surface using the MSMAll multimodal surface matching algorithm \citep{robinson2018multimodal}, yielding time series at 59,412 cortical vertices across both hemispheres and improving anatomical correspondence across subjects \citep{anticevic2008comparing, oosterhof2011comparison}. Further denoising included ICA-FIX artifact removal and nuisance regression of motion parameters, white matter, and cerebrospinal fluid signals.

We analyzed a final sample of 567 participants with complete rs-fMRI data from the first study visit (V1) and complete information on the selected phenotypic variables. Only V1 observations were included to avoid repeated measurements from participants who completed a follow-up visit (V2). Phenotypic variables were drawn from the HCP-A release and processed to remove variables with more than 50\% missing data, zero-variance columns, and collinear predictors ($|r| > 0.70$ or VIF $> 10$), yielding a final set of 12 covariates. Baseline characteristics of the study sample are summarized in Table~\ref{tab:baseline}.

\begin{table}[!htbp]
    \centering
    \begin{tabular}{ll}
        \toprule
        \textbf{Characteristic} & \textbf{Summary} \\
        \midrule
        Age, years & $61.1 \pm 15.3$ (36--89) \\[4pt]
        \textbf{Sex} & \\
        \quad Female & 323 (57.0\%) \\
        \quad Male   & 244 (43.0\%) \\[4pt]
        \textbf{Race} & \\
        \quad White                        & 426 (75.1\%) \\
        \quad Black or African American    & 64 (11.3\%)  \\
        \quad Asian                        & 40 (7.1\%)   \\
        \quad More than one race           & 22 (3.9\%)   \\
        \quad Other/unknown                & 15 (2.7\%)   \\[4pt]
        \textbf{Ethnicity} & \\
        \quad Hispanic or Latino           & 63 (11.1\%)  \\
        \quad Not Hispanic or Latino       & 503 (88.7\%) \\[4pt]
        Income, USD & 71,500 (40,000--110,000)$^*$ \\
        BMI, kg/m$^2$ & $27.2 \pm 4.9$ (16.6--45.9) \\
        Glucose & $101.9 \pm 16.4$ (42--298) \\
        PSQI global score & $4.84 \pm 2.99$ (0--17) \\[4pt]
        \textbf{Medication use} & \\
        \quad Blood pressure     & 164 (29.0\%) \\
        \quad SSRI/SNRI          & 44 (7.8\%)   \\
        \quad Cardiovascular     & 41 (7.2\%)   \\
        \quad Diabetes           & 21 (3.7\%)   \\
        \bottomrule
    \end{tabular}
    \caption{Baseline characteristics of the 567 HCP-Aging participants included in the analysis. Continuous variables are summarized as mean $\pm$ standard deviation (range); income is reported as median (interquartile range). Categorical variables are reported as count (percentage). $^*$Median (IQR).}
    \label{tab:baseline}
\end{table}

We selected covariates based on prior findings and the scientific aims of the study. Age and sex were included as primary demographic covariates. In our previous work on the ADHD-200 dataset, we identified significant associations between age and the long-memory parameter in children and adolescents \citep{Shahhosseini2025LM}. The Hurst exponent, which is linearly related to the long-memory parameter, has also been shown to vary systematically with age in resting-state fMRI \citep{dong2018hurst}, providing further motivation for including age as a covariate of interest. The HCP-A dataset extends this investigation to adults aged 36–89 years, allowing us to examine age-related associations with the long-memory parameter across a broader age range. Age was treated as a continuous variable and sex as binary.

Sleep quality was assessed using the global Pittsburgh Sleep Quality Index (PSQI) score \citep{joao2017validation}, a validated self-report measure of sleep disturbance, included as a continuous covariate. Poor sleep quality has been associated with reduced brain signal complexity and altered long-range temporal correlations in resting-state fMRI \citep{amorim2018poor}, motivating its inclusion. BMI and fasting glucose were included as continuous cardiometabolic markers. Prior work has linked vascular and metabolic factors to changes in brain signal scaling properties, with altered long-memory behavior observed in conditions involving cerebrovascular compromise \citep{maxim2005fractional}. Testosterone was included as a continuous hormonal covariate based on previous evidence of associations between sex hormones and functional brain connectivity and resting-state network dynamics during aging \citep{votinov2020effects}.

Four medication categories were also included as binary covariates: blood pressure medications, antidepressants (SSRI/SNRI), cardiovascular medications, and diabetes medications. Each category is associated with conditions that alter cerebrovascular function and resting-state brain dynamics in ways that may modulate the temporal structure of the BOLD signal. Hypertension and its treatment have been shown to affect neurovascular coupling and resting-state BOLD fluctuations, with blood pressure variability directly contributing to low-frequency BOLD signal variance \citep{whittaker2016beat}. Antidepressant treatment with SSRIs has been shown to alter resting-state functional connectivity between prefrontal and limbic regions \citep{anand2007reciprocal}, and depression itself is associated with changes in temporal variability of resting-state BOLD signals \citep{demirtacs2016dynamic}. Type 2 diabetes has been linked to accelerated brain ageing and reduced frontal fMRI activation \citep{antal2022type}, with glucose metabolism directly influencing the amplitude and frequency content of low-frequency BOLD fluctuations \citep{peters2020resting}. Including these medication categories therefore allows us to examine whether pharmacological status contributes independently to variation in long-memory structure across the cortical surface, and to account for medication-related differences when estimating associations with other covariates such as age.

Finally, memory and cognitive performance measures were included as continuous covariates to investigate whether the long-memory parameter of the fMRI BOLD signal is associated with cognitive functioning during aging. Previous studies have reported associations between temporal scaling properties of neural signals and memory-related cortical processing \citep{neufang2020regional, dong2018hurst}, supporting the inclusion of these variables.

For each subject, the resting-state fMRI time series, consisting of 1,912 time points at each cortical vertex, were transformed into the wavelet domain using a discrete wavelet transform (DWT).  We applied a second-order Daubechies wavelet transform with periodic boundary handling, yielding $T^* = 1{,}914$ wavelet coefficients per vertex across eight decomposition levels. The coefficient counts per level were 956, 478, 239, 120, 60, 30, 15, and 8, with an additional 8 approximation coefficients at the coarsest scale. Given the TR of 0.8 s, the eight decomposition levels correspond approximately to frequency bands spanning 0.31--0.63 Hz at the finest scale down to 0.005--0.010 Hz at the coarsest, covering the principal frequency range of resting-state BOLD fluctuations. Each vertex-level DWT coefficient $Z_{vk}$ was mean-centered prior to model fitting.

\subsection{Subject-Level Long-Memory Estimation}
Following the DWT, the long-memory parameter $\alpha_{s,v} \in (-1, 1)$ was estimated at each cortical vertex $v$ using the Bayesian hierarchical model described in Section~\ref{model}. The spatial CAR prior was defined on the cortical mesh graph, with adjacency determined by shared edges in the triangular mesh of the fsLR 32k surface, encouraging spatial smoothness in the estimated long-memory maps. For each subject, MCMC was run for 2,000 iterations, with the first 500 discarded as burn-in and the remaining samples thinned by a factor of two, yielding 750 posterior draws of $\alpha_{s,v}$ per vertex. 

\begin{figure*}[h]
\centerline{\includegraphics[width=\textwidth,height=0.32\textheight]{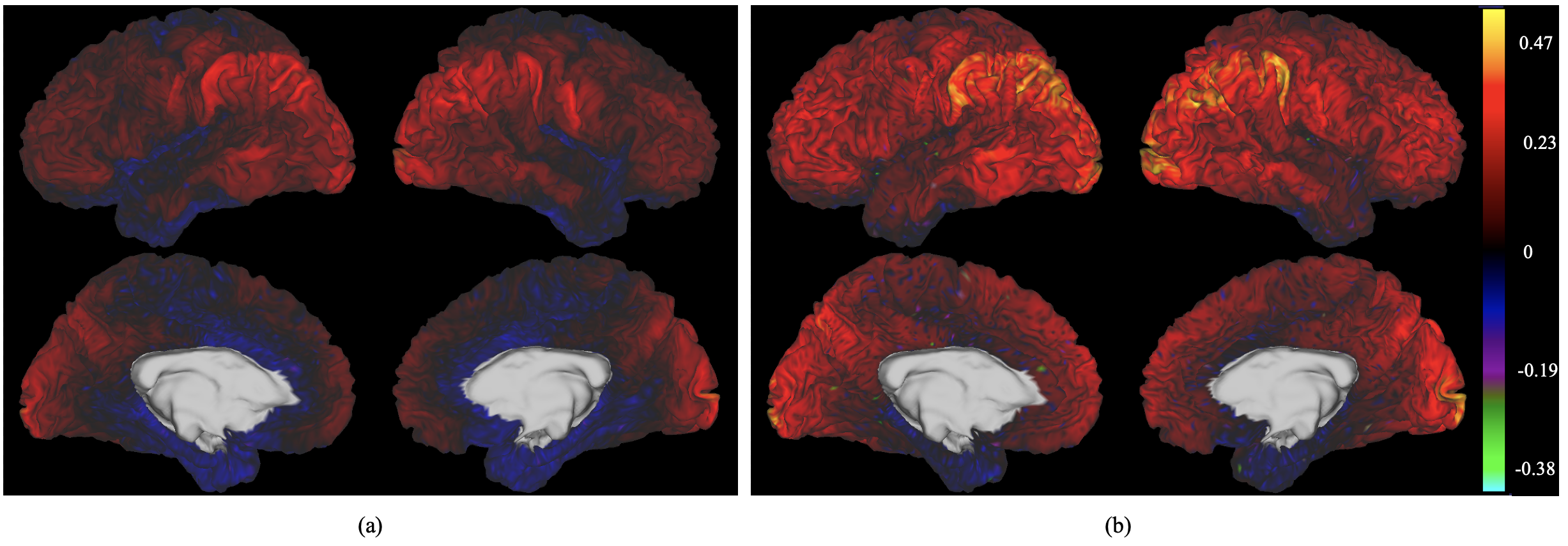}}
\caption{Group-averaged posterior mean of the long-memory parameter $\widehat{\alpha}_v$ mapped onto the cortical surface for two age subgroups: (a) 40 subjects aged 85 years and older, and (b) 66 subjects aged 40 years and younger. Both hemispheres are shown in lateral and medial views.\label{lm_map}}
\end{figure*}

To illustrate age-related differences in the spatial distribution of long-memory, Figure \ref{lm_map} presents the group-averaged posterior mean of the long-memory parameter for participants aged 40 years and younger (66 subjects) and those aged 85 years and older (40 subjects). The younger group shows higher positive long-memory values across broad frontal, parietal, and sensorimotor regions, with particularly elevated values in the right frontal and precentral cortex, extending into dorsal and medial prefrontal areas. This pattern is substantially attenuated in the older group, where corresponding regions show values closer to zero and a less pronounced spatial organization. Because larger positive values of the long-memory parameter indicate stronger temporal persistence and long-range dependence, these maps suggest lower temporal persistence in resting-state dynamics among older participants. This descriptive pattern is consistent with the negative age association identified in the Stage 2 analysis.

\subsection{Population-Level Inference}

To propagate Stage~1 posterior uncertainty into the group-level analysis, we adopted the nested MCMC strategy detailed in Section \ref{nested}. For each of the 750 posterior draws $\widehat{\alpha}_{s,v}^{(t)}$, $t = 1, \ldots, 750$, the full cross-subject matrix of long-memory estimates was assembled and used as the response in the Stage~2 Bayesian regression. This yielded 750 conditional Stage ~2 posterior distributions for the regression coefficients, each corresponding to a different realization from the Stage ~1 posterior. The final posterior was approximated by combining draws across all nested runs, producing ($750 \times 750 = 562{,}500$) posterior samples per vertex per covariate, so that Stage ~2 inference incorporates uncertainty propagated from Stage ~1.

The Stage~2 analysis examined associations between the long-memory parameter and 12 phenotypic covariates across the 567 participants, organized into a $567 \times 12$ design matrix. The covariates included age, sex, income, sleep quality (PSQI global score), BMI, fasting glucose, testosterone, blood pressure medications, antidepressants (SSRI/SNRI), cardiovascular medications, and diabetes medications, along with one memory and cognitive performance measure.

To address the high dimensionality of the cortical surface response, comprising 59,412 vertices, we applied the composite basis decomposition of \cite{miranda2023novel}. The long-memory maps were parcellated using the Schaefer atlas \citep{schaefer2018local}, and within each parcel, a local basis was constructed to capture 99\% of the within-parcel spatial variability. The long-memory maps were then projected onto these local bases, reducing the spatial dimension within each parcel. The resulting projections were combined across all 200 parcels, and a global basis was constructed to capture spatial patterns across parcels while retaining 99\% of the total variability.

Bayesian inference was performed in the reduced space. Posterior inference in the reduced-dimensional regression model was performed using Gibbs sampling, as the full conditional distributions of the regression coefficients $\boldsymbol{\beta}_r^*$ and variance parameters $\delta_r^2$ are available in closed form. The sampler was run for 2,000 iterations, with the first 500 discarded as burn-in and the remaining samples thinned by a factor of two, yielding 750 posterior draws per component. Each draw was projected back to the full cortical surface using the corresponding iteration-specific composite basis, recovering the vertex-level posterior distribution of the regression coefficients for each covariate.

To control for multiple comparisons across the 59,412 cortical vertices, we applied simultaneous credible bands (SimBaS) following the approach of \cite{ruppert2003semiparametric} to the posterior samples of $\beta_v$ obtained after projecting the reduced-space MCMC draws back to the full cortical surface. The key idea is to identify the smallest nominal level $\varsigma^*$ such that the $100(1-\varsigma^*)$\% simultaneous credible band excludes zero at a given vertex, and to declare that vertex significant if $\varsigma^* < 0.05$. Because this band is constructed jointly across all vertices, it provides simultaneous posterior  coverage over the  cortical surface. This makes SimBaS well suited to the high-dimensional and spatially structured inference problem encountered in cortical surface analysis.

\begin{figure*}[h]
\centerline{\includegraphics[width=0.6\textwidth]{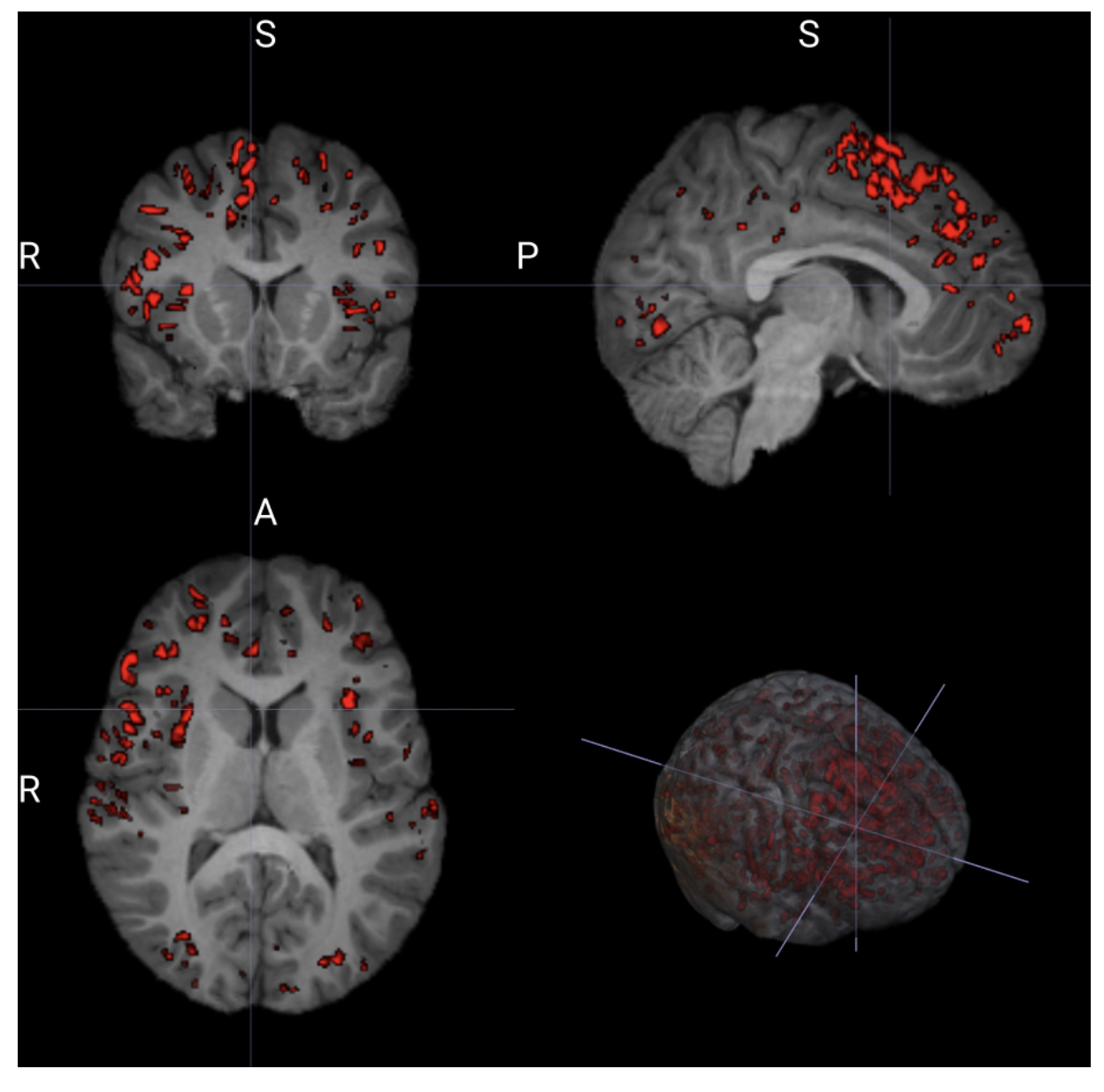}}
\caption{Significant age-related clusters in the long-memory parameter after SimBaS correction at $\varsigma^* < 0.05$, shown for the nested two-stage spatial Bayesian model.\label{age_binary}}
\end{figure*}

Of the 12 covariates considered, age was the only covariate significantly associated with the long-memory parameter after SimBaS correction at \(\varsigma^* < 0.05\). Figure \ref{age_binary} shows the significant vertices identified for the age effect. The corresponding binary significance map, provided in Appendix \ref{appendix.b}, contains numerous small and spatially isolated clusters across the cortical surface. To facilitate anatomical interpretation, we retained clusters containing at least 25 contiguous vertices for presentation in the main results. This threshold was guided by the observed cluster-size distribution. After applying this threshold, 18 clusters remained, ranging in size from 25 to 194 vertices. These clusters represent spatially coherent regions exhibiting significant age-related associations with the long-memory parameter.

\begin{figure*}[h]
\centerline{\includegraphics[width=0.6\textwidth]{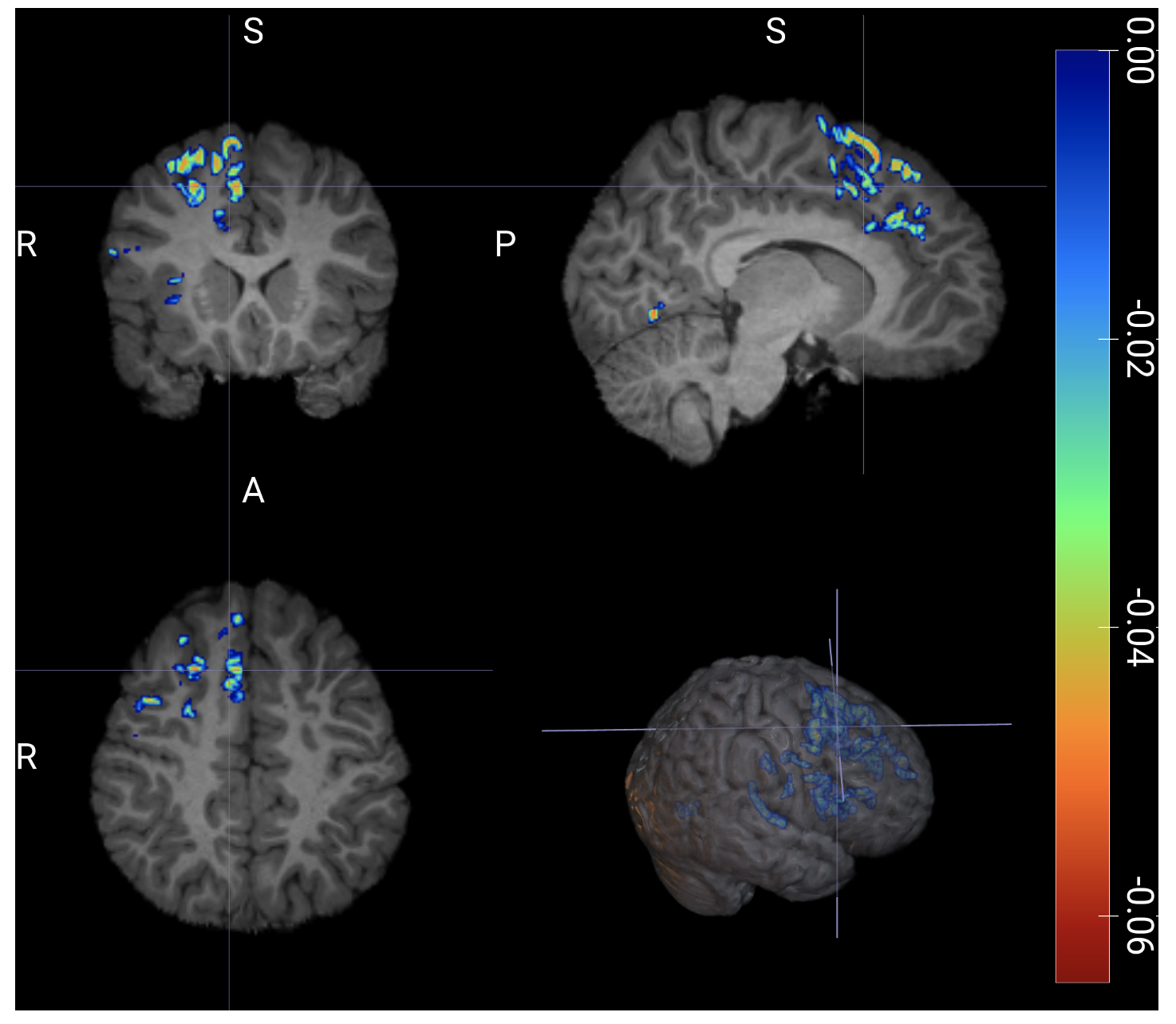}}
\caption{Posterior mean of the regression coefficient $\widehat{\beta}_v$ for the age covariate mapped onto the cortical surface, restricted to vertices belonging to significant clusters of 25 or more contiguous vertices after SimBaS correction at $\varsigma^* < 0.05$. Results are shown in coronal (upper left), sagittal (upper right), axial (lower left), and three-dimensional surface (lower right) views. All displayed vertices correspond to the proposed two-stage CAR model with nested MCMC.\label{coeff_map}}
\end{figure*}

Figure \ref{coeff_map} shows the posterior mean of the age regression coefficient $\widehat{\beta}_v$, restricted to significant clusters containing at least 25 contiguous vertices. All retained coefficients are negative, indicating that older age is associated with lower long-memory values within these regions. The significant associations are predominantly localized in the right hemisphere, particularly in the dorsal and medial prefrontal cortex, precentral cortex, and frontal eye fields. Significant clusters were identified within the Default Mode Network, including dorsal and medial prefrontal parcels 195 and 197; the Control Network, including medial prefrontal parcels 180 and 181; and the Dorsal Attention Network, including frontal eye field and precentral parcels 145–147. Together, the coefficient map shows that the negative age association is spatially localized rather than uniformly distributed across the cortical surface.

\begin{figure*}[h]
\centerline{\includegraphics[width=0.6\textwidth]{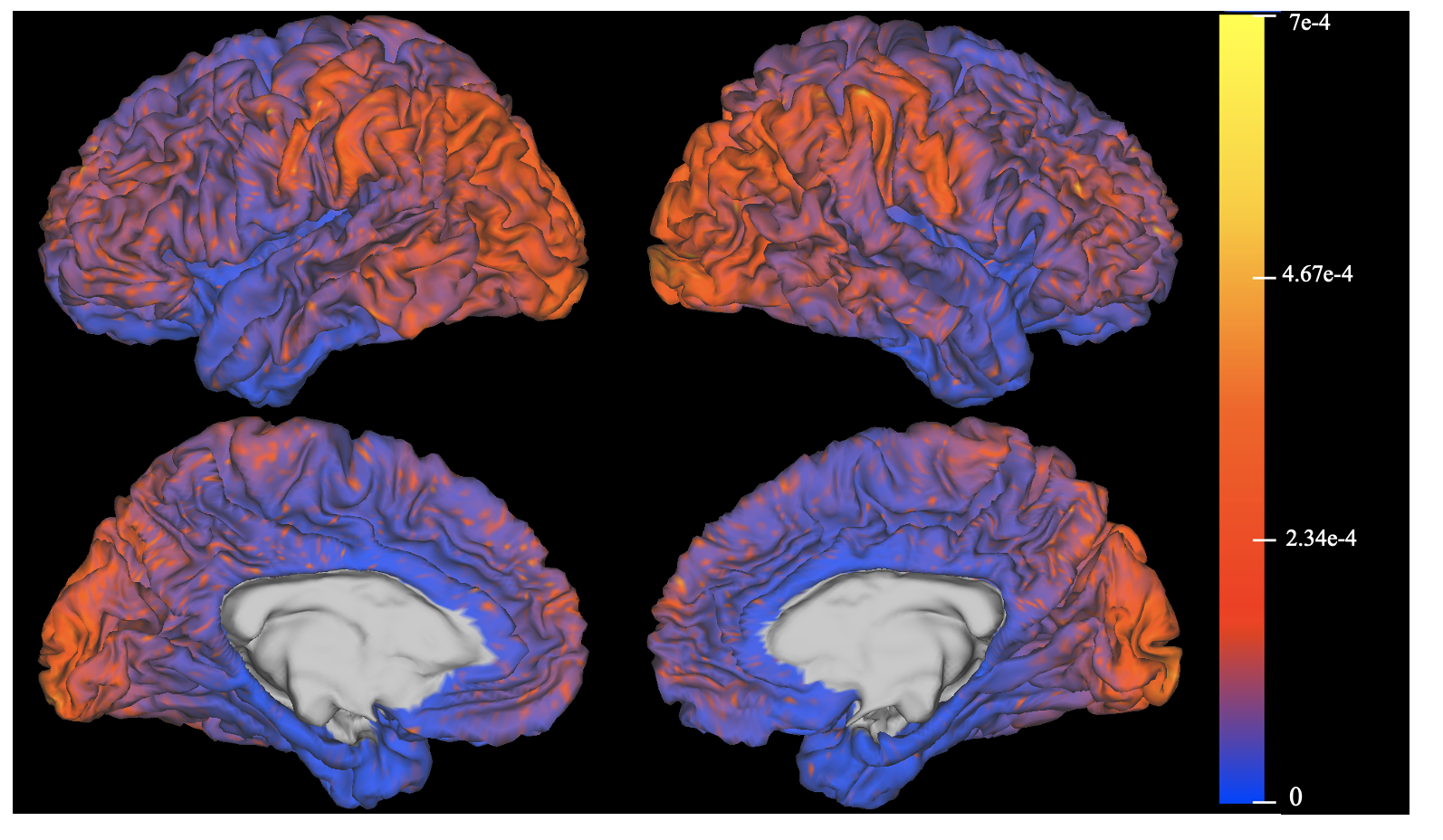}}
\caption{Posterior variance of the regression coefficient $\widehat{\beta}_v$ for the age covariate mapped onto the cortical surface. Results are shown in coronal (upper left), sagittal (upper right), axial (lower left), and three-dimensional surface (lower right) views. All displayed vertices correspond to the proposed two-stage CAR model with nested MCMC.\label{var_map}}
\end{figure*}

Figure \ref{var_map} presents the vertexwise posterior variance of the age regression coefficient under the proposed CAR nested model. These variances account for uncertainty in both subject-level LM estimation and population-level regression through nested MCMC.

\section{Discussion}\label{discussion_sec}

The age-related spatial pattern observed in the long-memory maps suggests that healthy aging is accompanied by a weakening of persistent, scale-free temporal organization in resting-state brain activity. The most pronounced differences were observed in frontal and precentral regions, where younger adults showed strongly positive long-memory values, whereas corresponding values in the oldest group were substantially reduced toward zero. This finding is consistent with previous studies reporting age-related reductions in the Hurst exponent, a related measure that is linearly related to the long-memory parameter, particularly in frontal and parietal regions \citep{dong2018hurst, churchill2016suppression, liu2013complexity}. From a multiscale perspective, the reduction in $\alpha$ reflects weaker long-range temporal dependence and is consistent with reports of reduced complexity at longer temporal scales with aging \citep{mcintosh2014spatiotemporal, sleimen2014aging, sleimen2015brain}. Together, these findings support the broader loss-of-complexity framework, suggesting that aging is associated with a reduced capacity to maintain coordinated dynamics across temporal scales.

Beyond the overall age-related reduction in long-memory dynamics, the spatial distribution of the regression coefficients indicates that negative age associations are particularly pronounced in frontal regions involved in cognitive control and attention. The concentration of negative effects in dorsal and medial prefrontal, frontal eye field, and precentral regions is consistent with \cite{dong2018hurst}, whose whole-brain analysis identified age-related decreases in the Hurst exponent predominantly in frontal regions, including the right middle frontal and precentral cortex. \cite{churchill2016suppression} similarly reported suppression of scale-free fMRI dynamics with healthy aging.

The predominance of frontal effects may also be viewed in the context of the frontal aging hypothesis, which proposes greater age-related vulnerability of prefrontal systems and their associated executive functions \citep{greenwood2000frontal}. Structural studies further indicate substantial age-related deterioration of frontal gray and white matter, providing a possible anatomical basis for altered temporal organization of activity in these regions \citep{cabeza2012frontal, petkoski2023white}. The negative associations observed across regions spanning the Default Mode, Control, and Dorsal Attention Networks may therefore reflect a weakening of persistent, long-range temporal dependencies in systems involved in executive function and attentional processing. This interpretation suggests a possible link between structural deterioration and changes in long-memory dynamics, but does not establish a causal relationship.

A key methodological contribution of the proposed framework is the incorporation of spatial dependence directly into the estimation of subject-specific cortical LM fields through the CAR prior. Rather than estimating LM parameters independently across cortical locations, the CAR prior allows information to be shared among neighbouring vertices according to the cortical surface topology, providing spatial regularization while preserving local variation. This is particularly important for vertexwise fMRI analysis, where independent estimation can produce spatially irregular or unstable estimates. By incorporating spatial dependence at Stage~1, the model uses the spatial organization of the cortical surface directly in LM estimation, rather than relying on spatial structure only at the subsequent dimension-reduction stage. The simulation results demonstrate the benefits of this approach: the CAR model substantially reduced subject-level LM estimation error, while both CAR-based approaches achieved lower population-level group-effect RMSE than the independent plug-in model across all simulated conditions. The CAR plug-in model also generally achieved higher spatial detection rates, supporting the value of spatial regularization for identifying localized group effects.

A second key methodological contribution is the nested MCMC approach, which propagates uncertainty from subject-specific LM estimation into population-level inference. Rather than conditioning on a single estimate of the LM field, nested MCMC integrates over its Stage 1 posterior distribution. By the law of total variance, this accounts for both the conditional variability of the regression coefficients $\boldsymbol{\beta}$ given the LM field $\boldsymbol{\alpha}$ and the additional variability arising from uncertainty in $\boldsymbol{\alpha}$. Under correct model specification, CAR nested and CAR plug-in produced nearly identical posterior mean estimates, with nested MCMC yielding slightly wider credible intervals. In the real-data analysis, the posterior variance maps under CAR nested characterize the spatial distribution of uncertainty in the estimated age effects. Together, these findings illustrate how the proposed framework incorporates Stage 1 estimation uncertainty into population-level inference.

Two directions may further extend the proposed framework. The first is the development of a fully joint Bayesian model in which subject-specific LM estimation and population-level inference are performed simultaneously, allowing information to be exchanged between the two levels. Such a formulation was computationally infeasible in the present study because of the high dimensionality of the cortical-surface data, with LM parameters estimated across tens of thousands of vertices for hundreds of subjects. More efficient inference and dimension-reduction strategies could make fully joint modeling feasible in future work.

The second direction is to extend the framework to volumetric fMRI data. This would require adapting the spatial prior to the three-dimensional neighborhood structure of voxels rather than the cortical-surface mesh. More flexible spatial priors could also be investigated to capture longer-range or region-based dependence beyond immediate spatial neighbors. Such priors could incorporate anatomical or functional relationships between brain regions, allowing correlated LM behavior across locations that are not spatially adjacent.

Overall, the proposed framework combines spatial modeling of subject-specific long-memory fields with uncertainty propagation in population-level Bayesian inference. The simulation results demonstrate the benefits of the CAR prior for estimation accuracy and spatial detection, while nested MCMC propagates subject-level uncertainty into population-level inference at a substantially lower computational cost than a fully joint MCMC approach. Applied to HCP-Aging resting-state fMRI data, the framework identified spatially localized negative associations between age and long-memory, particularly in frontal cortical regions.

\newpage
\bibliographystyle{plain}
\bibliography{wileyNJD-Harvard}

\newpage
\appendix
\section{Comparison of Posterior Variance Between Models}

\begin{figure*}[h]
\centerline{\includegraphics[width=\textwidth]{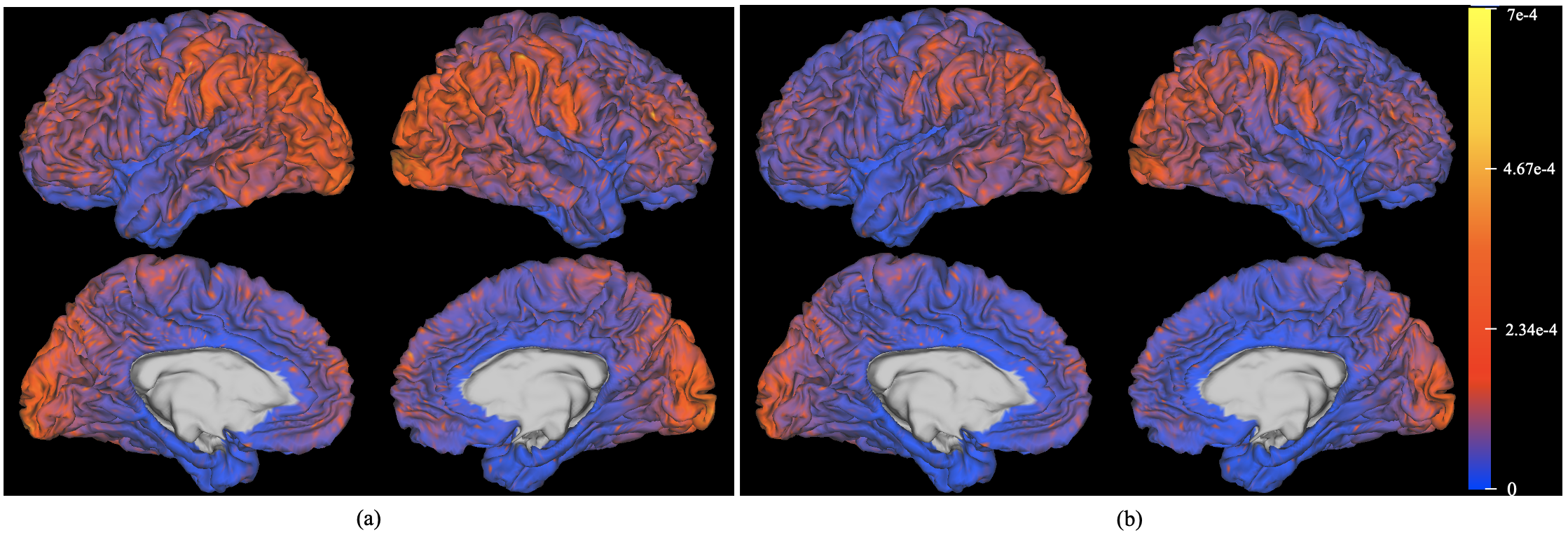}}
\caption{Spatial distribution of the posterior variance of the age regression coefficient across the cortical surface under the proposed model.\label{variance_map}}
\end{figure*}

Figure \ref{variance_map} compares the vertexwise posterior variance of the age regression coefficient under the proposed CAR nested model and the independent plug-in model. Both approaches exhibit similar spatial patterns, although posterior variances are generally larger under the CAR nested model.

\section{Posterior Inference}
This appendix provides the posterior derivations and computational updates used for inference under the two-stage Bayesian framework described in Section \ref{model}. The posterior computation follows the modular structure of the model. In Stage~1, subject-specific long-memory fields are estimated. In Stage~2, retained realizations from the subject-level posterior are projected onto the composite basis and incorporated into the population-level regression, for which the resulting conditional distributions are available in closed form and are sampled using Gibbs updates. The derivations below provide the posterior kernels and full conditional distributions used in each stage.
\subsection{Posterior inference for the two stage CAR model \label{appendixA.1}}
Posterior inference follows the two-stage structure of the model, with subject-level spatial estimation in Stage~1 and population-level regression in Stage~2.
\subsubsection*{Stage 1: Full conditional distributions}

This appendix presents the posterior derivations for the Stage 1 subject-level spatial CAR model. Each subject has its own parameters $\lambda_s$, $\nu_s$, $\sigma_s^2$, and $\kappa_s$. To simplify notation, the subject index $s$ is omitted throughout; all derivations apply independently to each subject. Recall that throughout this paper, $\alpha_v$ denotes the wavelet-domain long-memory parameter, represented through the unconstrained transformation
\begin{equation}
    \alpha_v = \tanh\!\left(\frac{\lambda_v}{2}\right),
\end{equation}
and that spatial dependence is introduced through the conditional autoregressive precision matrix
\begin{equation}
    Q(\kappa) = D - \kappa A.
\end{equation}
Let $Z_v = (Z_{v,1}, \dots, Z_{v,T^*})^\top$ denote the wavelet-domain observations associated with voxel $v$.

\subsubsection*{Joint Posterior Distribution}

For a single subject, the posterior distribution is proportional to
\begin{equation}
    p(\lambda, \nu, \sigma^2, \kappa \mid Z) \propto p(Z \mid \nu, \lambda)\, p(\nu)\, p(\lambda \mid \sigma^2, \kappa)\, p(\sigma^2)\, p(\kappa). \tag{A.1}
\end{equation}
The likelihood contribution follows from the wavelet-domain observation model,
\begin{equation}
    Z_{v,i} \mid \alpha_v, \nu_v \sim \mathcal{N}\!\left(0,\, \nu_v 2^{-m_i \alpha_v}\right),
\end{equation}
which implies the log-likelihood
\begin{equation}
    \log p(Z \mid \nu, \lambda) = -\frac{1}{2} \sum_{v=1}^{V} \left[ T^* \log \nu_v - (\log 2)\,\alpha_v M_1 + \frac{1}{\nu_v} \sum_{i=1}^{T^*} Z_{v,i}^2\, 2^{m_i \alpha_v} \right], \tag{A.2}
\end{equation}
where $M_1 = \sum_{i=1}^{T^*} m_i$. The CAR prior contributes
\begin{equation}
    \log p(\lambda \mid \sigma^2, \kappa) = \frac{1}{2}\log|Q(\kappa)| - \frac{V}{2}\log\sigma^2 - \frac{1}{2\sigma^2}\lambda^\top Q(\kappa)\lambda. \tag{A.3}
\end{equation}

\subsubsection*{Full Conditional Distribution of $\nu_v$}

We now derive the conditional posterior for the voxelwise innovation variance $\nu_v$. Since the remaining parameters are treated as fixed during this update, only terms involving $\nu_v$ are retained from the joint posterior.

For voxel $v$, the likelihood contribution is
\begin{equation}
    p(Z_v \mid \nu_v, \alpha_v) \propto \nu_v^{-T^*/2} \exp\!\left\{ -\frac{1}{2\nu_v} \sum_{i=1}^{T^*} Z_{v,i}^2\, 2^{m_i \alpha_v} \right\}. \tag{A.4}
\end{equation}
Define the scale-adjusted residual sum of squares
\begin{equation}
    S_v(\alpha_v) = \sum_{i=1}^{T^*} Z_{v,i}^2\, 2^{m_i \alpha_v}.
\end{equation}
Then the likelihood can be written compactly as
\begin{equation}
    p(Z_v \mid \nu_v, \alpha_v) \propto \nu_v^{-T^*/2} \exp\!\left\{ -\frac{S_v(\alpha_v)}{2\nu_v} \right\}.
\end{equation}
The prior specification is $\nu_v \sim \text{Inv-Gamma}(a_\nu, b_\nu)$, with kernel
\begin{equation}
    p(\nu_v) \propto \nu_v^{-(a_\nu+1)} \exp\!\left(-\frac{b_\nu}{\nu_v}\right).
\end{equation}
Combining likelihood and prior yields
\begin{equation}
    p(\nu_v \mid \text{rest}) \propto \nu_v^{-(a_\nu + T^*/2 + 1)} \exp\!\left\{ -\frac{b_\nu + \tfrac{1}{2}S_v(\alpha_v)}{\nu_v} \right\}. \tag{A.5}
\end{equation}
Recognizing this expression as the kernel of an inverse-gamma distribution gives
\begin{equation}
    \nu_v \mid \text{rest} \sim \text{Inv-Gamma}\!\left( a_\nu + \frac{T^*}{2},\; b_\nu + \frac{1}{2}S_v(\alpha_v) \right). \tag{A.6}
\end{equation}
This update is conditionally conjugate and can therefore be sampled directly within the Gibbs sampling scheme.

\subsubsection*{Full Conditional Distribution of $\sigma^2$}

We next derive the conditional posterior for the spatial variance parameter $\sigma^2$. Retaining only the terms involving $\sigma^2$ from the joint posterior in Equation~(A.1), the CAR prior contributes
\begin{equation}
    p(\lambda \mid \sigma^2, \kappa) \propto (\sigma^2)^{-V/2} \exp\!\left\{ -\frac{1}{2\sigma^2}\lambda^\top Q(\kappa)\lambda \right\}. \tag{A.7}
\end{equation}
The prior specification is $\sigma^2 \sim \text{Inv-Gamma}(a_\sigma, b_\sigma)$, with kernel
\begin{equation}
    p(\sigma^2) \propto (\sigma^2)^{-(a_\sigma+1)} \exp\!\left(-\frac{b_\sigma}{\sigma^2}\right).
\end{equation}
Combining likelihood and prior terms yields
\begin{equation}
    p(\sigma^2 \mid \text{rest}) \propto (\sigma^2)^{-(a_\sigma + V/2 + 1)} \exp\!\left\{ -\frac{b_\sigma + \tfrac{1}{2}\lambda^\top Q(\kappa)\lambda}{\sigma^2} \right\}. \tag{A.8}
\end{equation}
Recognizing this expression as the kernel of an inverse-gamma distribution gives
\begin{equation}
    \sigma^2 \mid \text{rest} \sim \text{Inv-Gamma}\!\left( a_\sigma + \frac{V}{2},\; b_\sigma + \frac{1}{2}\lambda^\top Q(\kappa)\lambda \right). \tag{A.9}
\end{equation}
This update is conditionally conjugate and can therefore be sampled directly within the Gibbs sampling procedure.

\subsubsection*{Full Conditional Distribution of $\kappa$}

The conditional posterior for the CAR dependence parameter $\kappa$ is obtained by retaining only the terms involving $\kappa$ from Equation~(A.1). Dependence on $\kappa$ enters through both the determinant and quadratic form associated with the CAR precision matrix. Combining these terms with the Beta prior yields
\begin{equation}
    p(\kappa \mid \lambda, \sigma^2) \propto |Q(\kappa)|^{1/2} \exp\!\left\{ -\frac{1}{2\sigma^2}\lambda^\top Q(\kappa)\lambda \right\} \kappa^{a_\kappa - 1}(1-\kappa)^{b_\kappa - 1}. \tag{A.10}
\end{equation}
Because both $|Q(\kappa)|$ and $\lambda^\top Q(\kappa)\lambda$ depend nonlinearly on $\kappa$, this conditional distribution is not available in closed form. Consequently, $\kappa$ is updated using a Metropolis--Hastings step.

To improve numerical stability and enforce support on $(0,1)$, sampling is performed on the logit scale,
\begin{equation}
    \zeta = \log\!\left(\frac{\kappa}{1-\kappa}\right).
\end{equation}
A random-walk proposal is defined as
\begin{equation}
    \zeta' = \zeta + \epsilon, \qquad \epsilon \sim \mathcal{N}(0, s_\kappa^2),
\end{equation}
where $s_\kappa$ is a proposal variance tuned during warm-up. The proposed value $\kappa' = \mathrm{logit}^{-1}(\zeta')$ is accepted with probability $\min(1, r)$, where the log acceptance ratio is
\begin{equation}
    \begin{aligned}
        \log r &= \frac{1}{2}\log|Q(\kappa')| - \frac{1}{2}\log|Q(\kappa)| - \frac{1}{2\sigma^2}\left[ \lambda^\top Q(\kappa')\lambda - \lambda^\top Q(\kappa)\lambda \right] \\
        &\quad + (a_\kappa - 1)\left(\log\kappa' - \log\kappa\right) + (b_\kappa - 1)\left[\log(1-\kappa') - \log(1-\kappa)\right] \\
        &\quad + \log\{\kappa'(1-\kappa')\} - \log\{\kappa(1-\kappa)\}.
    \end{aligned} \tag{A.11}
\end{equation}
The final term corresponds to the Jacobian adjustment induced by the logit transformation.

\subsubsection*{Full Conditional Distribution of $\lambda$ and HMC Gradient}

The transformed spatial parameter vector $\lambda = (\lambda_1, \dots, \lambda_V)^\top$ does not admit a conjugate full conditional due to the nonlinear relationship between $\lambda_v$ and the long-memory parameter $\alpha_v$. Consequently, posterior inference for $\lambda$ is performed using Hamiltonian Monte Carlo (HMC).

Retaining all terms involving $\lambda$, the conditional log-posterior kernel is
\begin{equation}
    \log \pi(\lambda) = \sum_{v=1}^{V} \left[ \frac{\log 2}{2}\alpha_v M_1 - \frac{1}{2\nu_v} S_v(\alpha_v) \right] - \frac{1}{2\sigma^2}\lambda^\top Q(\kappa)\lambda, \tag{A.12}
\end{equation}
where $\alpha_v = \tanh\!\left(\frac{\lambda_v}{2}\right)$. Efficient implementation of HMC requires the gradient of the log-posterior with respect to each component $\lambda_v$. Applying the chain rule yields
\begin{equation}
    \frac{\partial \log \pi}{\partial \lambda_v} = \frac{\partial \ell}{\partial \alpha_v} \cdot \frac{d\alpha_v}{d\lambda_v} - \frac{[Q(\kappa)\lambda]_v}{\sigma^2}. \tag{A.13}
\end{equation}
Differentiating the likelihood term with respect to $\alpha_v$ gives
\begin{equation}
    \frac{\partial \ell}{\partial \alpha_v} = \frac{\log 2}{2} M_1 - \frac{\log 2}{2\nu_v} \sum_{i=1}^{T^*} m_i Z_{v,i}^2\, 2^{m_i \alpha_v}. \tag{A.14}
\end{equation}
The derivative of the transformation $\alpha_v = \tanh\!\left(\frac{\lambda_v}{2}\right)$ is
\begin{equation}
    \frac{d\alpha_v}{d\lambda_v} = \frac{1 - \alpha_v^2}{2}. \tag{A.15}
\end{equation}
Substituting Equations~(A.14) and~(A.15) into Equation~(A.13) yields the gradient contribution for voxel $v$. Stacking these derivatives across voxels produces the full gradient vector $\nabla_\lambda \log \pi(\lambda)$, which is supplied to the leapfrog integrator within the HMC algorithm.

The quadratic term $Q(\kappa)\lambda$ is computed using sparse matrix operations, enabling efficient evaluation despite the high dimensionality of the imaging domain.

\subsubsection*{Stage 2: Full Conditional Distributions}

For a fixed retained Stage~1 iteration $t$, the Stage~2 updates are derived conditionally on the corresponding LM realization. To simplify notation, the nested-iteration superscript $t$ is suppressed throughout this subsection. Following Stage~1, the corresponding LM realization is incorporated into the group-level regression model described in Section~2.2.

Let $\alpha^* \in \mathbb{R}^{S \times PC_f}$ denote the matrix of subject-level LM maps projected onto the composite basis $\Upsilon$, and let $X \in \mathbb{R}^{S \times P}$ represent the design matrix of subject-level covariates. The reduced-space regression model is
\begin{equation}
    \alpha^* = X\beta^* + \varepsilon^*, \qquad \varepsilon_r^* \sim \mathcal{N}(0, \delta_r^2 I), \qquad r = 1, \dots, PC_f. \tag{A.16}
\end{equation}
Because the model factorizes across basis components, posterior updates may be derived independently for each component $r$. Let $\alpha_r^* \in \mathbb{R}^S$ denote the $r$-th column of $\alpha^*$, and let $\beta_r^*$ denote the corresponding regression coefficient vector.

\subsubsection*{Joint Posterior for Basis Component $r$}

For a fixed basis component $r$, the likelihood is
\begin{equation}
    \alpha_r^* \mid X, \beta_r^*, \delta_r^2 \sim \mathcal{N}(X\beta_r^*, \delta_r^2 I),
\end{equation}
which implies the density
\begin{equation}
    p(\alpha_r^* \mid X, \beta_r^*, \delta_r^2) \propto (\delta_r^2)^{-S/2} \exp\!\left\{ -\frac{1}{2\delta_r^2}(\alpha_r^* - X\beta_r^*)^\top(\alpha_r^* - X\beta_r^*) \right\}. \tag{A.17}
\end{equation}
The prior for the regression coefficients is specified as $\beta_r^* \mid \delta_r^2 \sim \mathcal{N}(\mu_0, \delta_r^2 \Lambda_0)$, with density kernel
\begin{equation}
    p(\boldsymbol{\beta}_r^* \mid \delta_r^2) \propto (\delta_r^2)^{-P/2} |\Lambda_0|^{-1/2} \exp\!\left\{ -\frac{1}{2\delta_r^2}(\boldsymbol{\beta}_r^* - \boldsymbol{\mu}_0)^\top \Lambda_0^{-1}(\boldsymbol{\beta}_r^* - \boldsymbol{\mu}_0) \right\}
\end{equation}
The residual variance is assigned an inverse-gamma prior $\delta_r^2 \sim \text{Inv-Gamma}(k, l)$, with kernel
\begin{equation}
    p(\delta_r^2) \propto (\delta_r^2)^{-(k+1)} \exp\!\left(-\frac{l}{\delta_r^2}\right). \tag{A.19}
\end{equation}

\subsubsection*{Full Conditional Distribution of $\beta_r^*$}

To derive the posterior update for $\beta_r^*$, we retain terms involving $\beta_r^*$ from the likelihood and prior. Combining Equations~(A.17) and~(A.18) yields
\begin{equation}
    \log p(\beta_r^* \mid \alpha_r^*, \delta_r^2) \propto -\frac{1}{2\delta_r^2}\left[ (\alpha_r^* - X\beta_r^*)^\top(\alpha_r^* - X\beta_r^*) + (\beta_r^* - \mu_0)^\top \Lambda_0^{-1}(\beta_r^* - \mu_0) \right].
\end{equation}
Expanding quadratic terms and collecting coefficients in $\beta_r^*$ gives
\begin{equation}
    \log p(\boldsymbol{\beta}_r^* \mid \boldsymbol{\alpha}_r^*, \delta_r^2) \propto -\frac{1}{2\delta_r^2}\left[ \boldsymbol{\beta}_r^{*\top}\!\left(X^\top X + \Lambda_0^{-1}\right)\boldsymbol{\beta}_r^* - 2\boldsymbol{\beta}_r^{*\top}\!\left(X^\top \boldsymbol{\alpha}_r^* + \Lambda_0^{-1}\boldsymbol{\mu}_0\right) \right]
\end{equation}

This expression corresponds to a multivariate Gaussian kernel. Therefore, the full conditional distribution is
\begin{equation}
    \beta_r^* \mid \alpha_r^*, \delta_r^2 \sim \mathcal{N}(\mu_n, \delta_r^2\Lambda_n), \tag{A.21}
\end{equation}
where the posterior covariance and mean are
\begin{equation}
    \Lambda_n = \left({X^\top X} + \Lambda_0^{-1}\right)^{-1},
\end{equation}
and
\begin{equation}
    \mu_n = \Lambda_n\left({X^\top \alpha_r^*} + \Lambda_0^{-1}\mu_0\right). \tag{A.22}
\end{equation}
Under the prior specification adopted in Section~2.2, this posterior update corresponds to a regularized least-squares estimator in basis space, with shrinkage governed jointly by the prior covariance and residual variance parameter.

\subsubsection*{Full Conditional Distribution of $\delta_r^2$}
We next derive the conditional posterior for the residual variance associated with basis component $r$. Retaining all terms involving $\delta_r^2$ from the likelihood, the conditional prior for $\boldsymbol{\beta}_r^*$, and the inverse-gamma prior for $\delta_r^2$ gives the corresponding full conditional distribution. The likelihood contribution is
\begin{equation}
    p(\boldsymbol{\alpha}_r^* \mid X, \boldsymbol{\beta}_r^*, \delta_r^2) \propto (\delta_r^2)^{-S/2} \exp\!\left\{ -\frac{1}{2\delta_r^2}(\boldsymbol{\alpha}_r^* - X\boldsymbol{\beta}_r^*)^\top(\boldsymbol{\alpha}_r^* - X\boldsymbol{\beta}_r^*) \right\}. \tag{A.23}
\end{equation}
Define the residual sum of squares for basis component $r$ as
\begin{equation}
    \mathrm{RSS}_r = (\boldsymbol{\alpha}_r^* - X\boldsymbol{\beta}_r^*)^\top(\boldsymbol{\alpha}_r^* - X\boldsymbol{\beta}_r^*).
\end{equation}
Because the conditional prior for $\boldsymbol{\beta}_r^*$ is also scaled by $\delta_r^2$, it contributes the quadratic term
\begin{equation}
    Q_r = (\boldsymbol{\beta}_r^* - \boldsymbol{\mu}_0)^\top \Lambda_0^{-1}(\boldsymbol{\beta}_r^* - \boldsymbol{\mu}_0).
\end{equation}
Combining the likelihood, the conditional prior for $\boldsymbol{\beta}_r^*$, and the inverse-gamma prior $\delta_r^2 \sim \operatorname{Inv\text{-}Gamma}(k,l)$ gives
\begin{equation}
    p(\delta_r^2 \mid \boldsymbol{\alpha}_r^*, \boldsymbol{\beta}_r^*) \propto (\delta_r^2)^{-\left(k + \frac{S+P}{2} + 1\right)} \exp\!\left\{ -\frac{l + \frac{1}{2}(\mathrm{RSS}_r + Q_r)}{\delta_r^2} \right\}. \tag{A.24}
\end{equation}
This expression corresponds to the kernel of an inverse-gamma distribution. Therefore,
\begin{equation}
    \delta_r^2 \mid \boldsymbol{\alpha}_r^*, \boldsymbol{\beta}_r^* \sim \operatorname{Inv\text{-}Gamma}\!\left( k + \frac{S+P}{2},\; l + \frac{1}{2}(\mathrm{RSS}_r + Q_r) \right). \tag{A.25}
\end{equation}
This update is conditionally conjugate and can therefore be sampled directly within the Gibbs sampler.
\subsubsection*{Gibbs Sampling Scheme for Stage 2}

Because both full conditional distributions derived above are available in closed form, posterior inference for the Stage 2 model proceeds through Gibbs sampling. For each retained basis component $r = 1, \dots, PC_f$, the sampler alternates between updates of $\beta_r^*$ and $\delta_r^2$, using the conditional distributions in Equations~(A.21) and~(A.25).

After posterior samples are obtained in the reduced basis space, regression effects are mapped back to voxel space through the composite basis representation. For covariate $p$, the voxelwise regression coefficient map at iteration $t$ is reconstructed as
\begin{equation}
    \boldsymbol{\beta}_p^{(t)} = \Upsilon\, \boldsymbol{\beta}_p^{*\,(t)} \in \mathbb{R}^V, \qquad p = 1, \dots, P.
\end{equation}
where $\boldsymbol{\Upsilon}$ denotes the composite basis matrix and $\boldsymbol{\beta}_p^{*(t)}$ is the sampled coefficient vector in basis space.

This reconstruction yields posterior samples of spatially regularized voxelwise regression effects over the imaging domain. Subsequent inference is performed using posterior summaries computed from these reconstructed coefficient maps.

\section{Raw Binary Map}
\label{appendix.b}

\begin{figure*}[h]
\centerline{\includegraphics[width=\textwidth]{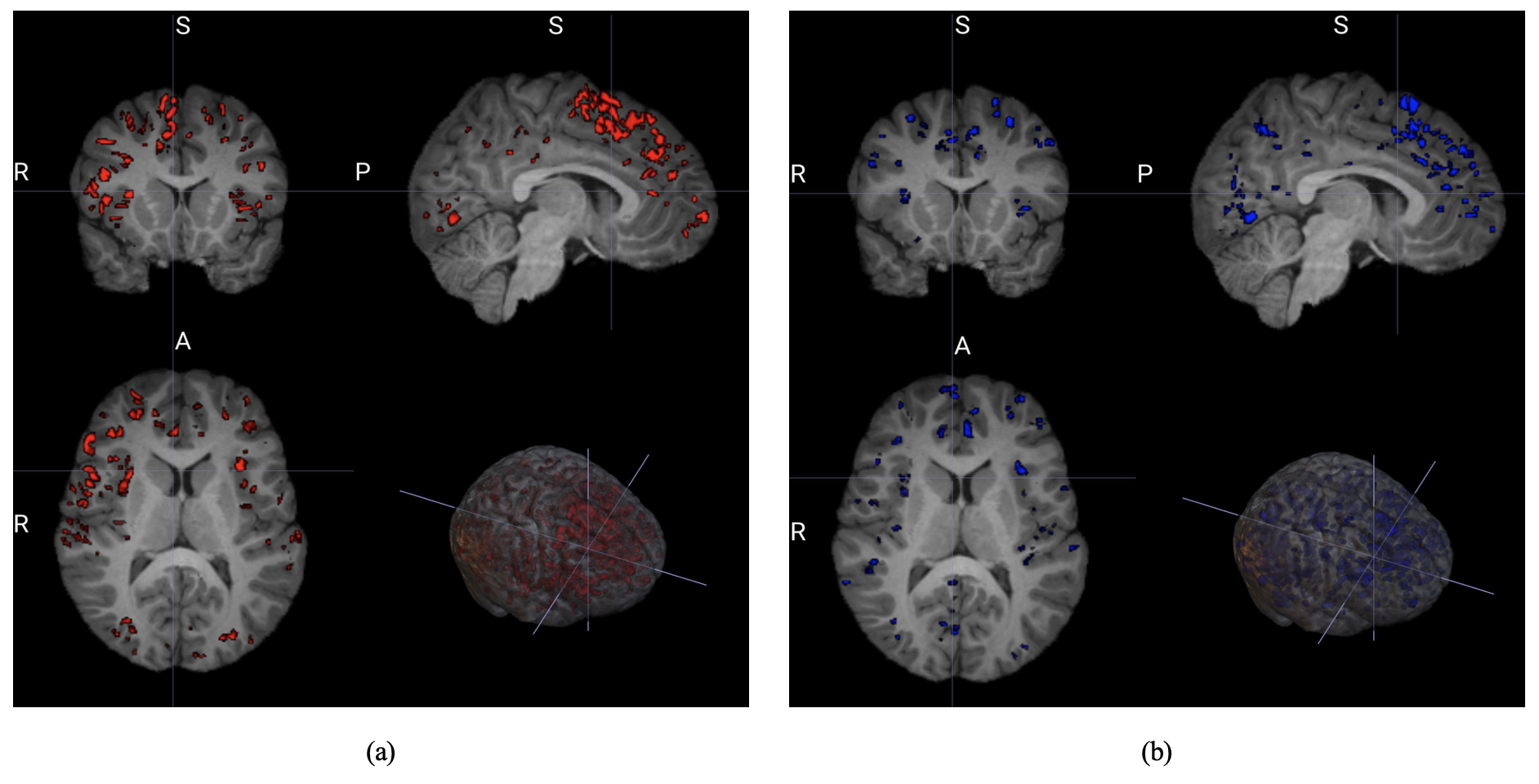}}
\caption{Raw binary map of significant vertices for the age covariate after SimBaS multiple-comparison correction at $\varsigma^* < 0.05$. (a) shows significant vertices identified by the nested two-stage spatial Bayesian model (red), and (b) shows significant vertices identified by the plug-in two-stage Bayesian model (blue).\label{age_raw_binary}}
\end{figure*}

\end{document}